\documentclass[sn-mathphys,Numbered]{sn-jnl}% Math and Physical Sciences Reference Style
\usepackage{graphicx}%
\usepackage{multirow}%
\usepackage{amsmath,amssymb,amsfonts}%
\usepackage{amsthm}%
\usepackage{mathrsfs}%
\usepackage[title]{appendix}%
\usepackage{xcolor}%
\usepackage{textcomp}%
\usepackage{manyfoot}%
\usepackage{booktabs}%
\usepackage{algorithm}%
\usepackage{algorithmicx}%
\usepackage{algpseudocode}%
\usepackage{listings}%
\usepackage{comment}
\usepackage{subfig}
\theoremstyle{thmstyleone}%
\newtheorem{theorem}{Theorem}%  meant for continuous numbers
\newtheorem{lemma}{Lemma}%
\theoremstyle{thmstyletwo}%

\theoremstyle{thmstylethree}%
\newtheorem{definition}{Definition}%

\newcommand{\mc}{\mathcal}
\newcommand{\mb}{\mathbf}

\begin{document}

\title[Quantifying the spread of communicable diseases with immigration of infectious individuals]{Quantifying the spread of communicable diseases with immigration of infectious individuals}


\author[1]{\fnm{Sof\'ia} \sur{Guarello}}\email{sofia.guarello@sansano.usm.cl}
\equalcont{These authors contributed equally to this work.}

\author*[1]{\fnm{Pablo} \sur{Aguirre}}\email{pablo.aguirre@usm.cl}
\equalcont{These authors contributed equally to this work.}

\author[1]{\fnm{Isabel} \sur{Flores}}\email{isabel.flores@usm.cl}
\equalcont{These authors contributed equally to this work.}

\affil*[1]{\orgdiv{Departamento de Matem\'atica}, \orgname{Universidad T{\'e}cnica Federico Santa Mar{\'i}a}, \orgaddress{\street{Avenida Espa\~na 1680}, \city{Valpara\'iso}, \postcode{Casilla 110-V}, %\state{State}, 
\country{Chile}}}

%\affil[2]{\orgdiv{Department}, \orgname{Organization}, \orgaddress{\street{Street}, \city{City}, \postcode{10587}, \state{State}, \country{Country}}}

%\affil[3]{\orgdiv{Department}, \orgname{Organization}, \orgaddress{\street{Street}, \city{City}, \postcode{610101}, \state{State}, \country{Country}}}


\abstract{ We construct a set of new epidemiological thresholds to address the general problem of spreading and containment of a disease with influx of infected individuals when the classic $\mathcal R_0$ is no longer meaningful.  We provide analytical properties of these indices and illustrate their usefulness in a compartmental model of COVID-19 with data taken from Chile showing a good predictive potential when contrasted with the recorded disease behaviour. This approach and the associated analytical and numerical results allow us to quantify the severity of an immigration of infectious individuals into a community, and identification of the key parameters that are capable of changing or reversing the spread of an infectious disease in specific models.}




\maketitle

\section{Introduction}\label{sec:intro}

The basic reproduction number ($\mc R_0$) is one of the most important indices in an epidemiological model. It indicates whether a disease is controlled or if it is spreading on the way to becoming an epidemic. The basic reproduction number $ \mc R_0 $ can be interpreted as the average number of secondary infections produced by a single primary infected individual in a fully susceptible population~\cite{brauer-cc, delamater, diekmann-book,murray, smith}. In other words, $\mc  R_0 $ is a measure of how many contagions are directly produced by the ``patient zero" of the disease. The basic reproduction number  is determined in a way that plays the role of a ``threshold'' for the spread of the disease. If $\mc  R_0> 1 $, more than one secondary infection occurs from the first infected case, and therefore, an epidemic is generated; on the other hand, if $ \mc R_0 <1 $, the disease is extinguished and the epidemic is prevented.

One of the most common techniques to calculate $\mc  R_0$ is based on the so-called Next Generation Matrix (NGM)~\cite{diekmann,MPG,vander}. This method has proven to be very versatile and general enough to be applied in different models in a variety of diseases \cite{brauer-cc, smith}.  However, the validity of the method is the same as that of a formal mathematical theorem: first it is necessary to verify that all the hypotheses and assumptions on which it was built are satisfied. Other more general techniques are also available
which (just like the NGM method) are based on the linearisation of the model equations around a {\it disease-free equilibrium point} and the analysis of its asymptotic stability~\cite{diekmann-book,smith}. Thus, this disease-free equilibrium is asymptotically stable if $\mc R_0<1$, and unstable if $\mc R_0>1$. %The main technical assumptions for a wide family of methods are the following~\cite{diekmann-book,diekmann,MPG,smith}:
%\begin{itemize}
%\item[(C1)] The disease-free set  is invariant. That is, for every initial condition located in this set, its solution remains confined in it as time moves forward.
%\item[(C2)] The disease-free set contains a {disease-free equilibrium} which also corresponds to an equilibrium of the full system.
%\end{itemize}

%If condition (C1) is not fulfilled at any point, this automatically implies that the model does not have a disease-free equilibrium and (C2) is also violated. In particular, it is not possible to find $\mc R_0$ based on the stability of said (non-existing) point nor to apply the NGM method to determine $\mc R_0$.

However, it is not possible to calculate $\mc  R_0 $ all the time. 
Such is the scenario, for example, in epidemiological models that include an influx of infectious individuals into the system from
the outside~\cite{Brauer,guo12, guo11,PE_1,Modelo_3,naresh,PE_2,tumwiine}. 
While a version of $\mc R_0$ may still be found ``in the limit" when there is no immigration of infected individuals, it loses validity as soon as the model allows entry of the immigrants into the system.
While a few works have attempted to deal with this loss of practical interpretability of~$\mc R_0$
%in models with incoming infected individuals
~\cite{Almarashi,Brauer,mccluskey,mclure}, (to the best of our knowledge) there is still the need to define a useful threshold to address general questions of propagation and containment of a disease in these scenarios and assess its value in issuing early warnings to decision-makers.

This work aims to understand  the influence of immigration of infected individuals for the propagation of a given transmissible disease, and to assess public health policies to mitigate the impact of such disease.
More specifically, we seek to propose and investigate new techniques to define an index that plays
a role analogous to that of the basic reproduction number $\mc R_0 $ (i.e., a threshold that determines spread or containment of a disease) in scenarios where  there is an inflow of infectious individuals into the system from
the outside. Such is the case, for example, in models for the spread of COVID-19~\cite{ayana,Modelo_inmigracion}, tuberculosis~\cite{guo11,Modelo_3}, AIDS/HIV~\cite{naresh}, malaria~\cite{tumwiine} and other communicable diseases in a country/state/city with open borders,
or STDs within prison population~\cite{carla2}, among other cases. The main challenges in these scenarios are:
\begin{itemize}
\item[(a)] It is not possible to calculate $\mc  R_0 $  by traditional methods; 
\item[(b)] And even if one could, the interpretation of such ``$\mc R_0$'' is (at best) very limited as an epidemiological threshold; see~\cite{Almarashi,Brauer,mclure}.
Currently, these problems pose massive challenges to decision makers, health workers, scientists, and ordinary people alike, and emerge as issues that have not yet been satisfactorily addressed in all their complexity~\cite{minsal,who2}.
\end{itemize}

Specifically, we will propose and study a set of indices ---framed in general, conceptual mathematical models--- which manage to capture quantitative aspects of growth and decay of a disease with immigration of infected people. This approach will allow us to isolate and identify common underlying mathematical ingredients in concrete systems with the aim of establishing sufficient conditions for the spread or containment of a particular disease. The advantage of this theoretical approach is that it allows one to find general features which might otherwise elude the researcher’s efforts when faced with a given particular model ---for instance, by remaining somewhat obscured by the specific details of the model equations. In this way, the benefit of establishing this mathematical formalism is that these findings are, hence, applicable in broader and applied contexts.

In order to convey the applicability and usefulness of our proposed methods, we consider a COVID-19 model fitted with data taken from Chile. Thus we show how our new indices can be used to take decisions that decrease the spread of the disease or avoid the increase of the propagation speed, in different scenarios. Our exploratory analysis results from modifying the different parameter values associated with either the epidemiological properties of the disease or hypothetical public health measures taken by Chilean authorities.
We also demonstrate how the analytical results obtained can be used to extract information on the long-term behavior of the COVID-19 disease.
In this way, we manage to pinpoint the key epidemiological parameters and their critical values that favor/prevent an epidemic outbreak in Chile. 

This paper is structured in the following way: Section~\ref{sec:method} presents the general setting and theoretical framework; the new propagation rates and thresholds are constructed and presented in section~\ref{sec:tasas} and illustrated in section~\ref{sec:modelo} with a model for COVID-19 with immigration of infectives and in section~\ref{sec:chile} with specific data taken from Chile; section~\ref{sec:asymptotic} showcases a study of asymptotic properties of the propagation rates; section~\ref{sec:discussion} summarises the main results and discusses open questions; and, finally, Appendix A briefly presents an additional way of constructing epidemiological thresholds similar to that introduced in the main part of this work.

%The second line of work seeks to apply these indices in concrete, meaningful models with immigration of infected individuals taken from both the relevant literature~\cite{brauer,elmojtaba,guo12, guo11,henshaw,hu,jia,naresh,sigdel,Modelo_inmigracion,tumwiine} and from collaboration with our research partners. 

 %%%%%%%%%%%%%%%%%%%%%%
\section{Mathematical framework}
\label{sec:method}
%%%%%%%%%%%%%%%%%%%%%%

 In a disease transmission model, individuals are placed into ``compartments'' based on a single discrete state variable. A compartment is called a disease compartment if the individuals there are infected; otherwise, it is a non-disease compartment. Suppose there are $n$ non-disease compartments and $m$ disease compartments, and let $\mb x\in\mathbb{R}^n$, $\mb y\in\mathbb{R}^m$ be the subpopulations in each of these compartments. Then, the compartmental model can be expressed in its general vector form
\begin{equation}\label{eq:general}
\left\{
\begin{array}{rcl}
\dfrac{d\mb x}{dt}&=& F(\mb x,\mb y), \vspace{2mm}\\
\dfrac{d\mb y}{dt}&= &  G(\mb x,\mb y). \\
\end{array}\right.
\end{equation}
Here, $(\mb x,\mb y)\in\mathbb{R}^n\times\mathbb{R}^m$ are the state variables as functions of time $t>0$ and we assume that $F:\mathbb{R}^n\times\mathbb{R}^m\rightarrow\mathbb{R}^n$ y $G:\mathbb{R}^n\times\mathbb{R}^m\rightarrow\mathbb{R}^m$ are $C^r$ functions, $r\geq1$. Here $\mb y $ groups the variables associated with disease states, while $ \mb x $ groups the disease-free state variables. 
%If $F=(f_1,\ldots,f_n)$ and $G=(g_1,\ldots,g_m)$, we may write \eqref{eq:general} as an equivalent system of $ n + m $ scalar ordinary differential equations
%
%\begin{equation}\label{eq:sist}
%\left\{
%\begin{array}{rcl}
%x_1' &=& f_1(x_1,\ldots,x_n,y_1,\ldots,y_m;\mu), \\
%&\vdots&\\
%x_n' &=& f_n(x_1,\ldots,x_n,y_1,\ldots,y_m;\mu), \\
%y_1' &=& g_1(x_1,\ldots,x_n,y_1,\ldots,y_m;\mu), \\
%&\vdots&\\
%y_m' &=& g_m(x_1,\ldots,x_n,y_1,\ldots,y_m;\mu), \\
%\end{array}\right.
%\end{equation}
%
%where $\mb x=(x_1,\ldots,x_n), \mb y=(y_1,\ldots,y_m).$

%Es decir, $\varphi_{p_0}$ satisface el problema de valor inicial $(d/dt)\varphi_{p_0}(t)=X(\varphi_{p_0}(t))$, para todo $t\in I$, y $\varphi_{p_0}(0)=p_0.$

The derivation of the basic reproduction number $\mc R_0$ by the NGM method and others is based on the linearisation of \eqref{eq:general} around a disease-free equilibrium point and the analysis of its asymptotic stability~\cite{diekmann,MPG}. Thus, this disease-free equilibrium is asymptotically stable if $\mc R_0 <1 $, and unstable if $\mc R_0> 1 $.
The main assumptions are as follows:
\begin{itemize}
\item[(C1)] The disease-free hyperplane 
$$\mc{S}=\{(\mb x,\mb y)\in\mathbb{R}^n\times\mathbb{R}^m:\ \ \mb y=\mb 0\in\mathbb{R}^m\}$$
is invariant under the flow generated by \eqref{eq:general}. That is, $G(\mb x,\mb 0)=0$, for every $(\mb x,\mb 0)\in \mc{S}$.
\item[(C2)] The disease-free subsystem $\mb x'=F(\mb x,\mb 0)$, $\mb x\in\mathbb{R}^n$, has an asymptotically stable equilibrium. 
(Here and in what follows, the prime denotes derivative with respect to time.)
% $y^*$.
\end{itemize}

If $\mc{S}$  is not invariant at any point ---and, hence, (C1) does not hold---, this automatically implies that (C2) is also violated. It follows that  \eqref{eq:general} does not have a disease-free equilibrium and it is not possible to apply the NGM method ---or any other method based on the existence of this equilibrium--- to determine $\mc R_0 $.
Such is the scenario, for example, in epidemiological models that include a positive flow of infectious individuals into the system from
the outside, i.e., our main focus of interest in this work.

Now, if  we add an immigration flow of individuals to each compartment and denote $F = (f_1 , ..., f_n)$ and $G = (g_1,...,g_m )$, we obtain a system of the form:
 \begin{equation}\label{sistema_general_1}
 \left\{
 \begin{array}{rcl}
         x_{1}' &= &\Pi p_1 +  f_{1}(x_1,...,x_n,y_1,...,y_m), \\
         & \vdots&\\
         x_{n}' &= &\Pi p_n + f_{n}(x_1,...,x_n,y_1,...,y_m),\\
         y_{1}' &= &\Pi p_{n+1} + g_{1}(x_1,...,x_n,y_1,...,y_m),\\
         & \vdots &\\
         y_{m} &= &\Pi p_{n+m} +  g_{m}(x_1,...,x_n,y_1,...,y_m),
 \end{array}
 \right.
    \end{equation} 
where $\Pi$ is the total immigration rate and $\textbf{p} = (p_1, ..., p_n, p_{n+1},...,p_{n+m})$ corresponds to the proportions of immigrants who enter each compartment and satisfy $\sum_{i=1}^{n+m}p_i = 1$. To ensure that the model includes immigration of infected individuals (whenever $\Pi$ is positive), we add the following assumption:\\

(C3) $p_k\geq 0$ for all $k=1,\ldots,n+m$, and $\displaystyle\sum_{i=1}^{m} p_{n+i}  > 0$.
  \\
  
It follows from (C1) and (C3) that \eqref{sistema_general_1} evaluated at a point $(\mb x,\mb 0) \in  \mathcal{S}$ satisfies
 \begin{equation}\label{S_invariante}
     \sum_{i=1}^{m} y_{i}' = \Pi  \sum_{i=1}^{m} p_{n+i} > 0.
 \end{equation}
As a result, the disease-free hyperplane $\mathcal{S}$ is not an invariant under \eqref{sistema_general_1}. Consequently, there is no disease-free equilibrium point for the system \eqref{sistema_general_1} when there is immigration of infected individuals from the outside.
Hence it is not possible to apply the NGM method (or any other method based on the existence of this equilibrium) to determine $\mathcal{R}_0$ ---and even if it were possible, it loses validity as an epidemiological index.
Even so, we want to measure how the appearance of a primary infected individual affects the evolution of the disease in a community initially composed of susceptible individuals.

%%%%%%%%%%%%%%%%%%%%%
\section{Propagation rates and epidemiological thresholds}
\label{sec:tasas}
%%%%%%%%%%%%%%%%%%%%%

Here we propose an alternative way to quantify the effect produced by the appearance of a patient zero in an initial disease-free scenario  in a model with immigation of infectives. %; that is, in the same spirit of the basic reproductive number but in cases when it is not possible to define $ R_0 $ by traditional methods.
In \eqref{sistema_general_1} there is no equilibrium contained in $\mc{S}$, i.e., there is no disease-free equilibrium.
We want to evaluate how the appearance of a primary infected individual affects the evolution of the disease in a community composed (initially) only of susceptible individuals. That is, we consider an initial state of \eqref{sistema_general_1} at a point of the form $\mb p_0=(\mb x ^ 0, \mb y ^ 0) \in \mathbb{R}^n\times \mathbb{R}^m $ such that:
\begin{enumerate}
\item All entries of $\mb x^0=(x^0_1,\cdots,x_n^0)\in\mathbb{R}^n$ are null except in that coordinate, say $x^0_1=N-1$, which corresponds to the susceptible compartment, where $ N $ is the total population at $t=0$.
\item All entries of $\mb y^0=(y^0_1,\cdots,y_m^0)\in\mathbb{R}^m$ are null except for a single coordinate, say $ y^0_i $, which corresponds to the appearance of an infectious individual in the state $ i $ of the disease. Namely, $y^0_i=1$, y $y^0_j=0$,  $j=1,\hdots,m$ with $j\neq i.$ 
\item $\mb{p_0}$ is not an equilibrium point of \eqref{sistema_general_1}.
\end{enumerate}

 Let us denote the vector field \eqref{sistema_general_1}  by $\mb X = \Pi \ \mb{p}+(F, G)^t $. 
The vector field $\mb X$ defines a flow which determines the solution vector
$$\varphi_{\mb p_0}(t)=(\varphi_1(t),\ldots,\varphi_{n+m}(t)):I\subset\mathbb{R}\rightarrow\mathbb{R}^n\times\mathbb{R}^m.$$
 Here $I$ is an interval of the real line containing $t=0$. Each entry of $\varphi_{\mb p_0}(t)$ expresses the value of the state variables at time $t$ starting from the initial condition $\varphi_{\mb p_0}(0)=\mb{p_0}$. Hence, 
$\varphi_{p_0}$ satisfies
\begin{equation}\label{eq:edo}
\frac{d}{dt}\varphi_{\mb p_0}(t)=\mb X(\varphi_{\mb p_0}(t)),
\end{equation}
 for every $t\in I\subset\mathbb{R}$. 
 
 %%%%%%%%%%%%%
\subsection{Main definitions}
%%%%%%%%%%%%%

Let $ \mb{u}\in\mathbb{R}^{n+m}$ be a unit vector orthogonal to $ \mc{S} $ with non-negative entries.
From \eqref{eq:edo}, the Euclidean inner product
\begin{equation}\label{eq:X.n}
\langle \mb X(\mb p_0), \mb u\rangle%|_{\mb{p_0}}
\end{equation}
is the component of the rate of change of the state variables in a direction of disease growth (i.e., orthogonal to $\mc S$) when the state of the system is distributed in its compartments according to $\mb p_0$. If the sign of \eqref{eq:X.n} is positive, the disease is spreading since the trajectory starting in $ \mb p_0 $ tends to move away from the disease-free hyperplane;
whereas if \eqref{eq:X.n} is negative, the disease is declining, as the solution gets closer to $\mc{S}$.

Once this growth or decrease of the disease has been determined by the sign of \eqref{eq:X.n}, we can obtain an instantaneous measure of this spread velocity  at $\mb p_0 $ relative to the change in the entire system. For this, we define the {\bf propagation rate at $p_0$ in the direction of u} as
\begin{equation}\label{eq:v}
X^{\mb u}(\mb{p_0})=\frac{\langle \mb X, \mb u\rangle}{\|\mb X\|}|_{\mb{p_0}},
\end{equation}
where $\|\cdot\|$ denotes the Euclidean norm in $\mathbb{R}^n\times\mathbb{R}^m$.
If $\mb X $ and $ \mb{u}$ are parallel vectors oriented in the same direction at  $ \mb p_0 $, then $X^{\mb u}(\mb{p_0}) = 1 $; thus, the solution of \eqref{sistema_general_1}  in $\mb p_0 $ is evolving in the direction of the disease growth at the highest possible rate. In turn, if $\mb X $ and $ \mb{u} $ are parallel vectors but in opposite directions, then $X^{\mb u}(\mb{p_0}) = -1 $; thus, the solution at $\mb p_0 $ is evolving in the direction of the decline of the disease at the highest possible rate.
Therefore, \eqref{eq:v} will have values in the range  $[-1,1]$, the extreme values $-1 $ and 1 being the maximum possible relative rates of decrease and spread, respectively. Note that \eqref{eq:v} keeps the sign of \eqref{eq:X.n}, since the norm $\|\mb X\|$ is always positive at $\mb{p_0}$, in such a way that if $X^{\mb u}(\mb{p_0}) <0 $, the disease declines, whereas if $X^{\mb u}(\mb{p_0})> 0 $ the disease spreads. In general, the smaller $ |X^{\mb u}(\mb{p_0})| $ is, the slower the spread or decline of the disease is at $\mb  p_0 $ in the direction of $ \mb u $. The case $ X^{\mb u}(\mb{p_0})= 0 $ represents a scenario in which the disease does not spread or decline in the direction of~$ \mb u $. Hence, condition 
$$ X^{\mb u}(\mb{p_0})= 0 $$ 
may be thought of as a critical or {\bf epidemiological threshold} value for the spread of the disease.

In general, a unit vector $\mb u$ orthogonal to $ \mc{S}$ will have the form
 \begin{equation}\label{eq:u-general}
 \mb u=(\underbrace{0,...,0}_\text{n},u_1,...,u_m),
 \end{equation}
where the coefficient vector $(u_1, ..., u_m)$ satisfies $\sum_{i=1}^mu_i^2=1$ and $u_i\geq0$, $i=1,\ldots, m$.
 %
 %$\mc{A}:=\left\{a_k\geq0,k=1,\ldots,n;\ \sum_{k=1}^na_k^2=1\right\}.$
 %
In this way, using the notation~\eqref{sistema_general_1}, the propagation rate \eqref{eq:v} can be expressed as
\begin{equation}\label{eq:v3}
    X^{\textbf{u}}(\mb p_0) =\frac{1}{\|\mb X(\mb p_0)\|}\displaystyle\sum_{i=1}^{m}u_{i} \big(\Pi\ p_{n+i} +\: g_{i}(\mb p_0)\big).
\end{equation}
%
%where $\| \mb{X}(\mb{p_0})\|=\sqrt{f_1^2(p_0)+\ldots+f_n^2(p_0)+g_1^2(p_0)+\ldots+g_m^2(p_0)}$.
{Note that  \eqref{eq:v3} may also be seen as a growth rate of aggregated (disease) variables in which the unit vector \eqref{eq:u-general} determines the weight assigned to each disease stage (or compartment). In concrete real case scenarios, these weights may be considered as costs associated with medical treatments, implementation of health policies, etc.}

Notice that formulas \eqref{eq:v} and \eqref{eq:v3}  can be readily extended to examine solutions at later stages of the evolution ---i.e., not just near $t=0$. More precisely, one may evaluate ${X}^{\mb u}$ at any arbitrary point $\mb p$ in the state space provided $\mb X(\mb p)\neq\mb{0}$ (i.e., $\mb{p}$ is not an equilibrium). In such case, the value and sign of ${X}^{\mb u}(\mb{p})$ indicate the growth tendency when the state of the system is distributed in its compartments according to $\mb{p}$.

%%%%%%%%%%%%%
\subsection{Particular cases of the propagation rate}
%%%%%%%%%%%%%

A particular case of  \eqref{eq:v3} occurs 
if the unit normal vector \eqref{eq:u-general} satisfies
$a_k=1,$ and  $a_i=0,$ for every $i\neq k$. Then \eqref{eq:v3} takes the form
\begin{equation}\label{eq:v4}
X^{\mb u}_k(\mb{p}_0):=\frac{1}{\| \mb X(\mb p_0)\|}\big(\Pi p_{n+k} + g_{k}(\mb p_0)\big).
\end{equation}
This formula corresponds to the relative rate of change of the solution in the direction of the $k$-th disease variable. In other words, both the sign and magnitude of  \eqref{eq:v4} quantify the contribution of the given initial state $\mb p_0$ (i.e., the appearance of the patient zero) to the increase in the number of individuals in the $k$-th disease stage. 

A similar interpretation may be given for different choices of~\eqref{eq:u-general}. For instance, another important case happens when $\mb u$ defines a uniform direction of increase of all (aggregated) disease variables, that is, 
 \begin{equation}\label{eq:u-uniforme}
 \textbf{u} = \frac{1}{\sqrt{m}}(\underbrace{0,...,0}_\text{n},\underbrace{1,...,1}_\text{m}).
 \end{equation}
 Then expression \eqref{eq:v3} reduces to
\begin{equation}\label{eq:v2}
{X}^{\mb u}(\mb{p}_0)=\frac{1}{\sqrt{m}\ \| \mb X(\mb p_0)\|}\sum_{i=1}^{m}\big(\Pi\ p_{n+i} + g_{i}(\mb p_0)\big).
\end{equation}
We call \eqref{eq:v2} the {\bf uniform propagation rate}.

%%%%%%%%%%%%%%%%%%%%%
\subsection{Propagation rates and thresholds in time}
\label{sec:tasas2}
%%%%%%%%%%%%%%%%%%%%%

The estimates given by  \eqref{eq:v}, \eqref{eq:v3}, \eqref{eq:v4} and  \eqref{eq:v2} are local at the initial point $ \mb p_0 $ and about the initial evolution of the solution, i.e., for $ 0 <t \ll 1 $.
 That is, they do not give information about the asymptotic (long-term) behavior of the system.
We can also investigate the evolution of ${X}^{\mb u}$ over time by letting the solution of  \eqref{eq:general} evolve from an arbitrary (regular) initial point $\mb p$ during a time interval of duration $ T> 0 $. In other words, we obtain a trajectory 
\begin{equation}\label{eq:tasa_tiempo}
X^{\mb u}_{\mb{p}}(t):=\left\{\frac{\langle \mb X, \mb u\rangle}{\|\mb X\|}|_{\varphi_{\mb{p}}(t)},\ 0\leq t\leq T\right\},
\end{equation}
constructed from points along the orbit segment $\left\{\varphi_{\mb{p}}(t),\ 0\leq t\leq T\right\}.$
Hence, trajectory $X^{\mb u}_{\mb{p}}(t)$ can be used to study the growth or decay of the disease in the direction of $\mb u$ along the entire ``history" of the disease that starts at $\mb{p}$ and up until~$t=T$. Furthermore, this allows us to
define a time average value of \eqref{eq:v} as % by letting the solution of  \eqref{eq:general} %evolve from the initial point $ p_0 $ during a time interval of duration $ T> 0 $, that is,
\begin{equation}\label{eq:promediot}
\langle X^{\mb u}_{\mb{p}}\rangle_T:=\frac{1}{T}\int_0^TX^{\mb u}_{\mb{p}}(t)\ dt=\frac{1}{T}\int_0^T\frac{\langle \mb X, \mb u\rangle}{\|\mb X\|}|_{\varphi_{\mb{p}}(t)}\ dt.
\end{equation}

Like  \eqref{eq:v} before,  \eqref{eq:promediot} will have values in $[-1,1]$ as well; the extreme values $-1 $ and 1 of $\langle X^{\mb u}_{\mb{p}}\rangle_T$ are the maximum possible relative rates of average decrease and spread, respectively. 
In particular, the sign of $\langle X^{\mb u}_{\mb{p}}\rangle_T$ tells us what the expected growth trend of the solution is, starting at $\mb p$ throughout its evolution (according to the model equations).  For instance, if $\langle X^{\mb u}_{\mb{p}}\rangle_T<0 $, the disease tends on average to decrease in the direction of $ \mb u $ rather than to increase over time. The estimate \eqref{eq:promediot} can be obtained for different choices of  \eqref{eq:u-general} giving information about the disease growth in the  direction of the given normal vector~$\mb u$. In particular, if $\mb u$ points in the direction of the $k$-th disease variable, from  \eqref{eq:v4}  we have
\begin{equation}\label{eq:promedio5}
\langle X^{k}_{\mb{p}}\rangle_T =\frac{1}{T} \int_0^T\frac{\Pi \ p_{n+k} + g_{k}(\mb p)}{\| \mb X(\varphi_{\mb p}(t))\|} dt.
\end{equation}
{This expression represents the average contribution of the initial state $\mb p$ to the variation of the $k$-th disease stage after a period $ T> 0 $}.

%%%%%%%%%%%%%%%%%%%%%%%%%%%%%%%%%%%%%%%
\section{Illustration: A model of COVID-19 with immigration of infectives}
\label{sec:modelo}
%%%%%%%%%%%%%%%%%%%%%%%%%%%%%%%%%%%%%%%

\begin{figure}[h!]
    \centering
    \includegraphics[width=\textwidth]{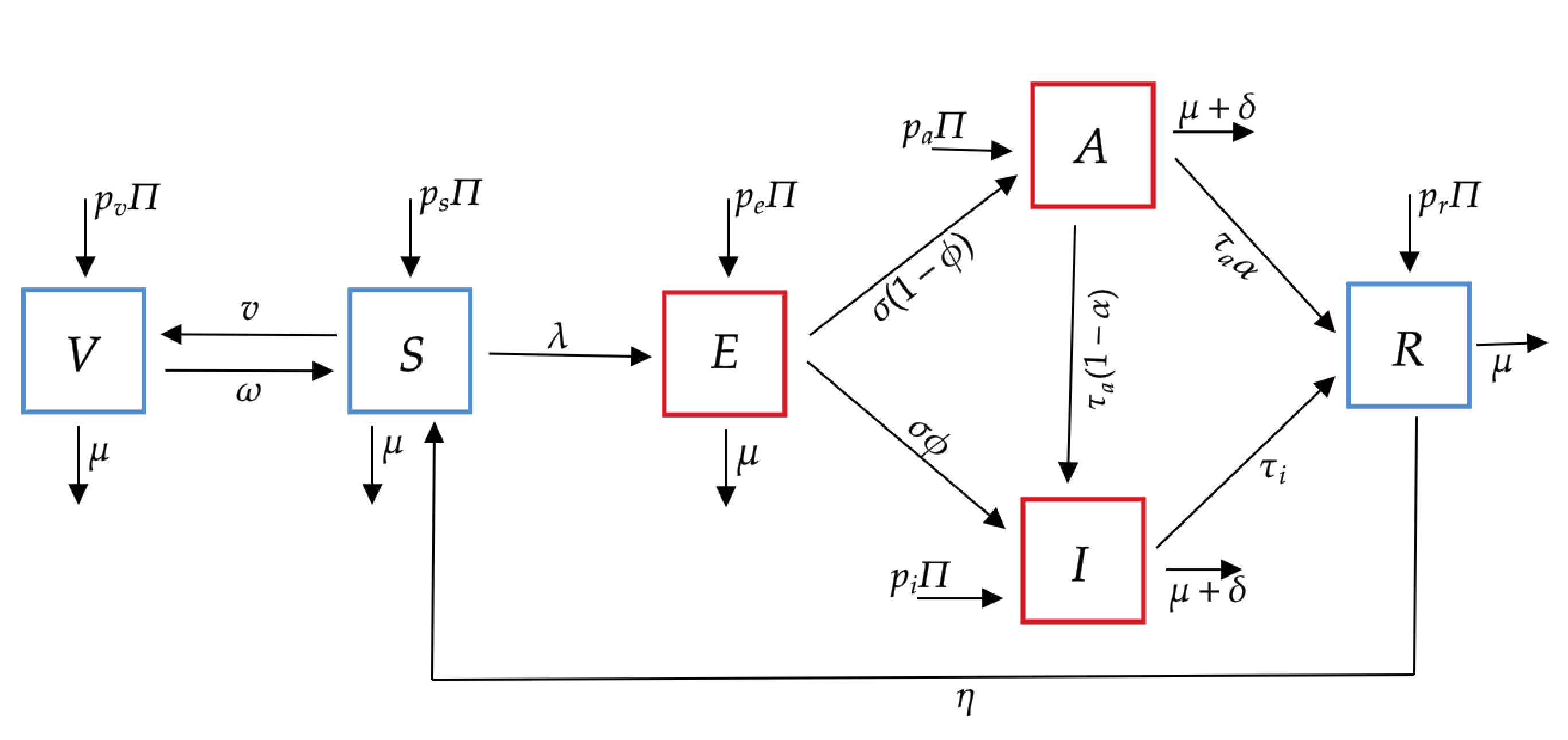}
    \caption{Diagram of the model of the transmission dynamics of COVID-19 with immigration of infected individuals~\eqref{sistema_SVEAIR}.}
    \label{fig:SVEAIR}
\end{figure}

In this section we will apply our proposed indices to study a model of COVID-19 with immigration of infected individuals taken from~\cite{Modelo_inmigracion}.
The population is divided into 6 compartments: Susceptible ($S$), Vaccinated ($V$), Recovered ($R$), Exposed ($E$), Asymptomatic ($A$), and Infected ($I$). The first three compartments correspond to non-infectious states and the other three correspond to different states associated with the disease. Figure \ref{fig:SVEAIR} shows a flow diagram of the model, from which the following system of nonlinear ordinary differential equations is derived (see details in~\cite{Modelo_inmigracion}):
\begin{equation}\label{sistema_SVEAIR}
    \begin{split}
    \begin{cases}
          &S' =  p_s \Pi + w V + \eta R - (\lambda + v  + \mu)S,\\
  &V' = p_v \Pi + v S -(w + \mu)V, \\
  &R' =  p_r \Pi  + \alpha\tau_a A + \tau_i I - (\mu + \eta)R,\\
   &E' =  p_e \Pi + \lambda S  - (\sigma + \mu )E, \\
   &I' =  p_i \Pi + \sigma \phi E + \tau_a (1-\alpha)A - (\tau_i + \mu + \delta)I,\\
   &A' =  p_a \Pi + \sigma (1- \phi) E   - ( \tau_a + \mu + \delta)A,
    \end{cases}
    \end{split}
\end{equation}
where $\lambda = \dfrac{\beta (\xi A + I)}{N}$ and $N=S+V+R+E+I+A$ is the total population. All the initial conditions in \eqref{sistema_SVEAIR} are taken to be non-negative. For the sake of simplification, we will define the parameter vector of the model as $\vartheta \in \mathbb{R}^{19}_{\geq0}$, where the subscript ``$\geq0$" indicates that all the parameters are non-negative.

The $S$ compartment corresponds to all those people who are susceptible to acquire the disease; $V$ corresponds to those individuals who were vaccinated and who have the immunity associated with the active vaccine; $R$ corresponds to those people who recovered from the disease and have active natural immunity; the Exposed compartment ($E$) corresponds to individuals who had contact with infected people with or without symptoms and are within the incubation period of the disease. Finally, the Infected ($I$) and Asymptomatic ($A$) compartments correspond to those infected with the disease with and without symptoms, respectively, whose diagnosis can be confirmed with some test. The model has a total rate of entry to the system given by $\Pi$.  The proportion of entry to each compartment is given by $p_s, p_v, p_r, p_e, p_i, p_a$, respectively, such that $p_s + p_v + p_r + p_e + p_i + p_a = $1. In particular, $p_e\Pi, p_i\Pi, p_a\Pi$ represent the rates of admission to the compartments associated with the disease, i.e, exposed, asymptomatic infected and symptomatic infected, respectively. Additionally, $\mu$ represents the natural mortality rate from each compartment.

The transmission flow of the disease is presented in figure \ref{fig:SVEAIR}. A susceptible person ($S$) can move to the exposed compartment ($E$) when being infected at a rate $\xi \beta$ by asymptomatic persons ($A$) and at a rate $\beta$ by infected ( $I$). Here $\beta$ represents the contact rate and $\xi$ represents the reduction in asymptomatic transmission. Total transmission is represented by $\lambda$; therefore, there is an entry to the exposed compartment at a rate of $\lambda$, which is caused by infections. The disease can evolve from $E$ to $I$ at a rate $\sigma \phi$, or to $A$ at a rate $\sigma(1-\phi)$.  Here $\sigma$ indicates the exit rate from $E$ or, seen in another way, $\sigma^{-1}$ indicates the average number of days that a person remains in the incubation period of the disease, while $\phi$ represents the proportion of exposed individuals who become infected with some symptom.  
The asymptomatic individuals ($A$) can evolve to be symptomatic compartment ($I$) at a rate $\tau_a (1-\alpha)$ or they can evolve to the recovery compartment ($R$) at a rate $\tau_a \alpha$, where $\tau_a$ represents the exit rate from compartment $A$ and $\alpha$ is the proportion of asymptomatic individuals who recover from the disease.
Similarly, the symptomatic infected ($I$) move to the recovered compartment ($R$) at a rate of $\tau_i$.
It should be noted that there is another exit rate from the infectious compartments ($I, A$) that is given by $\delta$, which represents the mortality rate associated with the disease. Once people enter the recovered compartment ($R$) they can re-enter the susceptible compartment ($S$) at a rate $\eta$, which represents the loss of natural immunity. Finally, susceptible people who are vaccinated enter at a rate $v$ to the $V$ compartment and remain there until they lose immunity at rate $\omega$.

%%%%%%%%%%%%%%%%%%%%%
\subsection{Preliminary results}
%%%%%%%%%%%%%%%%%%%%%

Due to the entry of infected persons into the system, model  \eqref{sistema_SVEAIR} has a single asymptotically stable endemic equilibrium point and the disease cannot be eradicated unless the immigration of infectives is stopped~\cite{Modelo_inmigracion}.
Hence, we carry out an analysis of the propagation rate defined in \eqref{eq:v} for  system \eqref{sistema_SVEAIR} in order to propose containment alternatives. In this section we limit ourselves to find analytical expressions for \eqref{eq:v} and leave their evaluations in real data for the following sections.

Let us consider an initial distribution of the population given by 
$$\mb p_0 = (S_0, V_0, R_0, E_0, I_0, A_0).$$
The vector field \eqref{sistema_SVEAIR} evaluated at $p_0$ is given by:
\begin{equation}
\label{ec:X(p_0)}
 \begin{split}
    \mathbf{X}(\mb p_0) =&  
    \bigg( p_s \Pi  + \omega V_0 + \eta R_0 - \left(\frac{\beta (\xi A_0 + I_0 )}{N_0} + v  + \mu\right) S_0,\\
    & p_v \Pi + v S_0 - (\omega + \mu) V_0, \\
     & p_r \Pi  + \alpha\tau_a A_0 + \tau_i I_0 - (\mu + \eta)R_0,\\
    & p_e \Pi + \frac{\beta (\xi A_0 + I_0 )}{N_0} S_0  - (\sigma + \mu )E_0,\\
   & p_i \Pi + \sigma \phi E_0 + \tau_a (1-\alpha)A_0 - (\tau_i + \mu + \delta)I_0,\\ 
   &p_a \Pi + \sigma (1- \phi)E_0   - (\tau_a + \mu + \delta)A_0 \bigg),
    \end{split}
\end{equation}
where $N_0 = S_0 + V_0 + R_0 + E_0 + I_0 + A_0 $.

Considering the spread of COVID-19 in the direction of growth of the exposed compartment $E$ category, we take the normal vector
$\textbf{u} = (0,0,0,1,0,0)$
and evaluate \eqref{eq:v} to obtain
\begin{equation}\label{sistema1_tasa_expuestos}
   X^{E}(\mb p_0)= \frac{ p_e \Pi +\beta (\xi A_0 + I_0 ) \dfrac{ S_0}{N_0}  - (\sigma + \mu )E_0}{\|\textbf{X}(\mb p_0)\|}.
\end{equation}
The value of $ X^{E}(\mb p_0)$ tells us if the disease is spreading or decreasing in the direction of $E$ and its rate of change in that direction. 
From \eqref{sistema1_tasa_expuestos} we can define a threshold manifold given by:
\begin{equation}\label{umbral_exp}
  \{ \vartheta \in \mathbb{R}^{19}_{\geq 0} | \ X^{E}(\mb p_0)=0\} \Leftrightarrow  \left\{ p_e \Pi + \beta (\xi A_0 + I_0 ) \dfrac{ S_0}{N_0}  - (\sigma + \mu) E_0 = 0 \right\}.
\end{equation}
It should be noted that \eqref{umbral_exp} can be interpreted as a threshold both in state space (or initial conditions) and in parameter space.
In the first case, for fixed parameter values, \eqref{umbral_exp} tells us the locus of all the points in state space where there is no increase or decrease of population in compartment $E$. Similarly, if the point $(S_0,V_0,R_0,E_0,I_0,A_0)$ is fixed, \eqref{umbral_exp} indicates the set in parameter space where the same property is satisfied. One could even consider a third option by taking the product space of the state variables $(S_0,V_0,R_0,E_0,I_0,A_0)$ and the vector of parameters $\vartheta$, obtaining from \eqref{umbral_exp} a threshold in such extended space.

On the other hand, taking  the entry proportion of those exposed to the system ($p_e$)  as a variable in \eqref{sistema1_tasa_expuestos} we obtain:
\begin{equation}\label{derivada_exp}
  \frac{\partial X^{E} }{\partial p_e} (\mb p_0) = \frac{\Pi \left( \| X(\mb p_0)\|^2 -  \left(p_e \Pi + \beta (\xi A_0 + I_0 ) \dfrac{ S_0}{N_0}  - g_1 E_0 \right)^2 \right) }{\|\textbf{X}(\mb p_0)\|^3} \geq 0.
\end{equation}
This indicates that regardless of the parameter values, the propagation rate of the disease in the direction of $E$ is increasing with respect to the entry proportion of the same type of infected individuals into the system. This is a consistency property one would expect to occur in any good propagation index.

Analogously, we now seek to measure the spread of COVID-19 from the infected ($I$) compartment. For this we consider
$\textbf{u} = (0,0,0,0,1,0),$
thus obtaining the rate of spread in the direction of the symptomatic infected as
\begin{equation}\label{sistema1_tasa_infectados}
   X^{I}(\mb p_0)= \frac{p_i \Pi + \sigma \phi E_0 + \tau_a (1-\alpha)A_0 - (\tau_i + \mu + \delta)I_0}{\|X(\mb p_0)\|}.
\end{equation}
Here, the threshold manifold is given by
\begin{equation}\label{umbral_inf}  \left\{ p_i \Pi + \sigma \phi E_0 + \tau_a (1-\alpha)A_0 - (\tau_i + \mu + \delta)I_0 = 0 \right\}.
\end{equation}
In addition, in this case there is also
\begin{equation}\label{derivada_inf}
  \frac{\partial X^{I}(\mb p_0) }{\partial p_i}  = \frac{\Pi \left( \|X(\mb p_0)\|^2 - ( p_i \Pi + \sigma \phi E_0 + \tau_a (1-\alpha)A_0 - g_{2}I_0)^2 \right) }{\|X(\mb p_0)\|^3} \geq 0.
\end{equation}
Thus, as expected, the propagation rate in the direction of the infected is  an increasing function of the proportion of entry of the infected into the system.

%Notar que en este caso particular, el conjunto umbral es un hiperplano en el espacio de estados, pues \eqref{umbral_inf} es lineal con respecto a $(E_0,A_0,I_0)$. En efecto, graficamos el plano \eqref{umbral_inf} en el espacio tridimensional $(E_0,A_0,I_0)$ utilizando los valores de parámetros de la tabla \ref{tabla_0} y obtuvimos la figura \ref{fig:umbral_inf}. Si el estado inicial de los compartimentos asociados a la enfermedad $(E_0,A_0,I_0)$ se encuentran en el plano $X^{I}(p_0)=0$ entonces inicialmente no hay crecimiento ni decrecimiento del compartimento $I$. Ahora, si el estado inicial se encuentran debajo del plano $X^{I}(p_0)=0$ entonces la tasa de propagación \eqref{sistema1_tasa_infectados} es positiva y los infectados están creciendo en ese instante. Por otro lado, si el estado inicial se encuentra por sobre el plano $X^{I}(p_0)=0$ entonces el compartimento de infectados está decreciendo en ese instante.

%\begin{figure}[t]
%    \centering
%    \includegraphics[scale=0.3]{Imagenes/diagram-20221121.png}
%    \caption{Gráfico del conjunto umbral \eqref{umbral_inf} en el espacio de estados $(E_0, A_0, I_0)$ utilizando los valores de parámetros de la Tabla \ref{tabla_0}}
%    \label{fig:umbral_inf}
%\end{figure}

Considering now the propagation in the  growth direction of the asymptomatic $A$ we take
$\textbf{u} = (0,0,0,0,0,1),$
thus obtaining the rate
\begin{equation}\label{sistema1_tasa_asintomática}
  X^{A}(\mb p_0)= \frac{p_a \Pi + \sigma (1- \phi)E_0   - (\tau_a + \mu + \delta)A_0}{\|X(\mb p_0)\|},
\end{equation}
which measures whether the disease is spreading or decreasing towards the asymptomatic infected $A$. In this case, the threshold manifold is given by:
\begin{equation}\label{umbral_asin}
     \left\{ p_a \Pi + \sigma (1- \phi)E_0   -(\tau_a + \mu + \delta) A_0 = 0 \right\}.
\end{equation}
The rate of spread in the direction of asymptomatic individuals $A$ increases with respect to the proportion of the same class of infected entering the system, namely,
\begin{equation}\label{derivada_asin}
  \frac{\partial X^{A}(\mb p_0) }{\partial p_a}  = \frac{\Pi \left( \|X(\mb p_0)\|^2 - ( p_a \Pi + \sigma (1- \phi)E_0   - g_3 A_0)^2 \right) }{\|X(\mb p_0)\|^3} \geq 0. 
\end{equation}

%que al igual que en dirección a los infectados, es un hiperplano que depende solo de los estados $E_0$ y $A_0$. Esto lo graficamos utilizando los valores de parámetros de la tabla \ref{tabla_0} obteniendo la recta en el plano $(E_0, A_0)$, ver figura \ref{fig:umbral_asin}. Luego, si el estado inicial de expuestos y asintomáticos $(E_0, A_0)$ está en la recta $ X^{A}(p_0)= 0$, entonces inicialmente el compartimento de los asintomáticos no está creciendo ni decreciendo. Ahora, si el estado inicial está por debajo de la recta $ X^{A}(p_0)= 0$ entonces la cantidad de asintomáticos está creciendo. Por el contrario, si el estado inicial está por sobre la recta $ X^{A}(p_0)= 0$, entonces el compartimento de asintomáticos está decreciendo en ese instante.

%\begin{figure}[t]
 %   \centering
%    \includegraphics[scale=0.25]{Imagenes/diagram-20221121-2.png}
 %   \caption{Gráfico del conjunto umbral \eqref{umbral_asin} } en el espacio de estados $(E_0,A_0)$ utilizando los valores de parámetros de la Tabla \ref{tabla_0}.
 %   \label{fig:umbral_asin}
%\end{figure}

Finally, we  measure the spread of COVID-19 in a uniform direction to the aggregated three states associated with the disease, that is, we consider
$\mb u = \frac{1}{\sqrt{3}}(0,0,0,1,1,1),$
to obtain
\begin{equation}\label{sistema1_tasa_unif}
\begin{array}{rcl}
    X^{U}(\mb p_0)&=& \dfrac{( p_e + p_i + p_a) \Pi +\beta (\xi A_0 + I_0 ) \dfrac{ S_0}{N_0}}{\sqrt{3}\ \|X(\mb p_0)\|}  \vspace{2mm}\\
      &&-  \dfrac{\mu (E_0 + I_0 + A_0) -  \delta (A_0 + I_0) - \tau_a \alpha A_0 - \tau_i I_0     }{\sqrt{3}\ \|X(\mb p_0)\|}.
      \end{array}
\end{equation}
It is interesting to note that the sign of \eqref{sistema1_tasa_unif} depends on all the state variables except the number of vaccinated people $V_0$.

%que indica si la enfermedad se está propagando o decreciendo en dirección uniforme a los tres estados infecciosos y su velocidad de crecimiento o decrecimiento en esa dirección. También se puede interpretar como una medida del crecimiento de la enfermedad si agrupamos los tres estados asociados a la enfermedad en un solo compartimento. 
In this case, the threshold manifold is given implicitly by equation:
\begin{equation}\label{umbral_unif}
   \begin{split}
     ( p_e + p_i + p_a) \Pi + \beta (\xi A_0 + I_0 ) \dfrac{ S_0}{N_0}   -  \mu (E_0 + I_0 + A_0) -  \delta (A_0 + I_0) - \tau_a \alpha A_0 - \tau_i I_0  = 0.
   \end{split} 
\end{equation}
Observe that if we take a small perturbation from \eqref{umbral_unif} in the form $p_k = p_k + \epsilon$ for any $k \in \{e,i,a\}$, with $\epsilon>0$ (resp. $<0$), then \eqref{sistema1_tasa_unif} moves to the region where $X^{U}(p_0) >0$ (resp. $<0$). In other words, after any small disturbance from the threshold manifold in any of the entry proportions associated with the disease, the spread of COVID-19 at that moment will start to increase or decrease depending on the sign of the perturbation.

Let us consider \eqref{sistema_SVEAIR}  with inmigration of infected, i.e., $p_e,  p_i, p_a > 0 $ and let us take a point $\mb p_0$ at the instant prior to the onset of the disease, i.e, with the entire population in the susceptible compartment:
$ \mb p_0 = (N_0, 0,0,0,0,0).$
From the formulas \eqref{sistema1_tasa_expuestos},  \eqref{sistema1_tasa_infectados}, \eqref{sistema1_tasa_asintomática} and \eqref{sistema1_tasa_unif}, we obtain:
\begin{equation}\label{eq:S=N}
\begin{array}{rclcrcl}
       X^{E}(\mb p_0) &=& \dfrac{ p_e \Pi   }{\|\mathbf{X}(\mb p_0) \|},&&
          X^{I}(\mb p_0) &=& \dfrac{p_i \Pi }{\|\mathbf{X}(\mb p_0) \|}, \vspace{2mm}\\
  X^{A}(\mb p_0) &=& \dfrac{p_a \Pi}{\|\mathbf{X}(\mb p_0) \|},\ &&
           X^{U}(\mb p_0) &=& \dfrac{( p_e + p_i + p_a) \Pi }{\|\mathbf{X}(\mb p_0) \|},
           \end{array}
           \end{equation}
where $ \mathbf{X}(\mb p_0) $ denotes expression \eqref{ec:X(p_0)} evaluated in $(N_0, 0,0,0,0,0)$, i.e.:
\begin{equation}
 \begin{split}
    \mathbf{X}(\mb p_0) =&  
    \big( p_s \Pi - ( v  + \mu)N_0, \: p_v \Pi + v N_0 , \:
     p_r \Pi,\:
     p_e \Pi,\:
    p_i \Pi,\:
   p_a \Pi \big),
    \end{split}
\end{equation}

It is easy to see that all the rates are positive due to the entry proportions of infected. In other words, in the particular case in which there is still no infected person but there is an influx of infected people into the system, the disease will inevitably spread within the system. This scenario can be interpreted as the moment in which the arrival of the first infected individual in the system is ``imminent'' and will produce an initial increase in the number of individuals at each stage of the disease.

Now, if we take the case in which there is no immigration of infected individuals, taking $p_e = p_i = p_a = 0$ in \eqref{eq:S=N} we obtain:
$X^{E}(\mb p_0) = X^{I}(\mb p_0) = X^{A}(\mb p_0) = X^{U}(\mb p_0) = 0.$
This result is consistent with what one would observe in reality. Indeed,  there is no infected person in the system and new ones are not yet entering; hence, the disease is neither increasing nor decreasing.
From the dynamics of \eqref{sistema_SVEAIR}, this is due to the fact that the initial point $\mb p_0$ is located in the disease-free hyperplane $\mc{S}$, which is an invariant hyperplane for the case without immigration of infected people.
Indeed, we have $E'=0, I'=0, A'=0$ in  \eqref{sistema_SVEAIR} at any point of the form
$ \mb p_0 = (S_0 , V_0 , R_0, 0,0,0)\in\mc{S}$
when $p_e = p_i = p_a = 0$.

%%%%%%%%%%%%%%%%%%%%%%%%%%%%
\subsection{More epidemiological thresholds in model \eqref{sistema_SVEAIR}}
%%%%%%%%%%%%%%%%%%%%%%%%%%%%

Using the method of the next generation matrix, the authors in \cite{Modelo_inmigracion} calculated the basic reproduction number $\mathcal{R}_0$ of \eqref{sistema_SVEAIR} in the case of no immigration of infected ($p_e=p_i=p_a =0 $), obtaining:
\begin{equation}\label{R0_SVEAIR}
    \mathcal{R}_0 = \frac{[(1-\alpha)(1-\phi)\tau_a + g_2 \xi (1-\phi ) + g_3 \phi ][\mu p_s] + \omega (p_s + p_v)]\sigma \beta}{g_1 g_2 g_3 (p_s+p_v)(\mu + v + \omega )},
\end{equation}
where $g_1 = \sigma + \mu$, $g_2 = \tau_{i} + \mu + \delta$ and $g_3 = \tau_a + \mu + \delta$. 

%mover umbrales manifolds desde seccion anterior.

Let us now consider \eqref{sistema_SVEAIR}  with inmigration of infected, i.e., $p_e,  p_i, p_a > 0 $ and 
 consider the scenario at the very start of the epidemics with a single infectious individual. Let $\mb p_0$ be a point in state space with a first infected person located in some compartment associated with the disease in an otherwise fully susceptible population. 

%%%%%%%%%%%%%%%%%%%%%%%%%%%%
\subsubsection{Patient zero in exposed state $E$}
%%%%%%%%%%%%%%%%%%%%%%%%%%%%

For instance, consider this first infected individual in the exposed compartment $E$, i.e.,
\begin{equation}\label{primer_expuesto}
    \mb p_0 = (S_0, 0,0,1,0,0),
\end{equation} 
where $S_0=N_0-1$ and $E_0=1$. 
From \eqref{sistema1_tasa_expuestos},  \eqref{sistema1_tasa_infectados}, \eqref{sistema1_tasa_asintomática} and \eqref{sistema1_tasa_unif} we have:
\begin{equation}\label{Tasa_1_expuesto_SVEAIR}
    \begin{split}
       X^{E}(\mb p_0) = \frac{p_e \Pi   - (\sigma + \mu ) }{\|{\textbf{X}(\mb p_0)}\|}, & \hspace{5mm}
         X^{I}(\mb p_0) = \frac{p_i \Pi + \sigma \phi }{\|{\textbf{X}(\mb p_0)}\|},\\
           X^{A}(\mb p_0)= \frac{ p_a \Pi + \sigma (1- \phi)}{\|{\textbf{X}(\mb p_0)}\|},& \hspace{5mm}
           X^{U}(\mb p_0)= \frac{( p_e + p_i + p_a) \Pi - \mu }{\|{\textbf{X}(\mb p_0)}\|},
    \end{split}
\end{equation}
%donde
%\begin{equation*}
%\begin{split}
%    \norm{\textbf{X}(p_0)}^2 =& ( p_s \Pi  - \left( v  + \mu\right) S_0)^2 + ( p_v \Pi + v S_0)^2+
%     ( p_r \Pi)^2+
 %   (  p_e \Pi   - (\sigma + \mu ))^2 +
%    (p_i \Pi + \sigma \phi)^2 +
%   (p_a \Pi + \sigma (1- \phi) )^2 
%\end{split}
%\end{equation*}
%
where  $\mathbf{X}(\mb p_0) $ corresponds to \eqref{ec:X(p_0)} evaluated in
\eqref{primer_expuesto}, that is:
\begin{equation}\nonumber
 \begin{array}{rcl}
    \mathbf{X}(\mb p_0) &=& 
    \Big( p_s \Pi  - \left( v  + \mu\right) S_0,\:
     p_v \Pi + v S_0 , \:
      p_r \Pi , \vspace{2mm} \\
     && p_e \Pi   - (\sigma + \mu ),\:
    p_i \Pi + \sigma \phi ,\:
  p_a \Pi + \sigma (1- \phi) \Big).
    \end{array}
\end{equation}

Assuming  $p_e, p_i, p_a > 0$ we have 
$           X^{I}(\mb p_0)  > 0$ and
            $X^{A}(\mb p_0) > 0$
for all parameter values. Then, the $I$ and $A$ compartments will initially grow after the appearance of a first infected person in the exposed state. However,
$X^{E}(\mb p_0)   < 0$ if and only if $p_e \Pi   - (\sigma + \mu ) < 0.$
Hence, 
the propagation rate in the direction of the exposed will be negative as long as the speed of entry into the exposed compartment ($p_e \Pi$) does not exceed the exit rate from the $E$ compartment due to deaths or evolution of the disease towards compartments $I$ and $A$ ($\mu + \sigma$). Equivalently,
$X^{E}(p_0)<0$ as long as
$
    \frac{1}{p_e \Pi} > \frac{1}{\sigma + \mu },
$
that is, if the expected time for a new individual in stage $E$ to enter the system is greater than the average time they remain in compartment $E$.

If we now consider the growth in a uniform direction to the three infectious states, from \eqref{Tasa_1_expuesto_SVEAIR} it is easy to see that:
\begin{equation*}
        X^{U}(\mb p_0)  < 0  \Leftrightarrow    \frac{( p_e + p_i + p_a) \Pi}{\mu}    < 1. 
\end{equation*}
This suggests one to define a number analogous to $\mathcal{R}_{0}$ in the following form:
\begin{equation}\label{R_E0}
    \mathcal{R}_{E_0} = \frac{( p_e + p_i + p_a) \Pi}{\mu}, 
\end{equation}
where the subscript $E_0$ indicates that the first infected individual is in the exposed state. The number \eqref{R_E0}  indicates the evolution of the disease at an initial instant of time with a first infected in the exposed state in a fully susceptible population {\em and} with entry of infected individuals into the system.
The threshold value of $  \mathcal{R}_{E_0}$ is 1:
If $ \mathcal{R}_{E_0} > 1$ the disease tends to increase, and if $ \mathcal{R}_{E_0} < 1$ it tends to decrease. Therefore, with a first exposed infected one has
\begin{equation*}
   X^{U}(\mb p_0) < 0 \Leftrightarrow \mathcal{R}_{E_0} < 1.
\end{equation*}

Note that if the proportions of infected people entering the country ($p_e, p_i, p_a$) or the number of people entering the country ($\Pi$) are increased, the value of $\mathcal{R}_{E_0} $ also increases. On the other hand, if the natural mortality rate ($\mu$) is increased, the value of $\mathcal{R}_{E_0}$ decreases (Increasing the mortality rate means that individuals ``exit'' the system more quickly). In other words, if the first infected individual is in the exposed state, the growth of the disease will be contained if the net rate of entry of infectious individuals (in all its stages) is less than the natural mortality rate.

%En particular,  considerando el caso \textbf{sin} inmigración de infectados ($p_e = p_i = p_a = 0$), de \eqref{Tasa_1_expuesto_SVEAIR} se obtiene que el signo de las tasas son,
%\begin{equation*}
%    \begin{split}
%        X^{E}(p_0) &< 0,\\
%           X^{I}(p_0) & > 0,\\
%            X^{A}(p_0) & > 0,\\
%             X^{U}(p_0) & < 0.
%    \end{split}
%\end{equation*}
%Es decir, ante un primer infectado en estado expuesto y en ausencia de inmigración de infectados, los expuestos van a decaer inicialmente, los infectados y asintomáticos a crecer y, si consideramos los tres estados infecciosos con un peso uniforme, en promedio la enfermedad va a decrecer. Resultado que tiene sentido pues, el primer y único individuo enfermo en estado expuesto no contagia a otras personas aún y solo puede evolucionar a corto plazo a estar infectado sintomático o  asintomático. Además, no ingresan nuevos infectados del exterior. De hecho, de \eqref{R_E0} se tiene
%$$\mathcal{R}_{E_0} = 0,$$
%el cual es menor que 1.

%%%%%%%%%%%%%%%%%%%%%%%%%%%%
\subsubsection{Patient zero in symptomatic infected state $I$}
%%%%%%%%%%%%%%%%%%%%%%%%%%%%

If the first infected 
%
%\begin{equation}\label{primer_infectado}
   $ \mb p_0 = (S_0, 0,0,0,1,0)$
%\end{equation} 
%
is in the symptomatic state $I$
we obtain
\begin{equation}\label{Tasa_1_infectado_SVEAIR}
    \begin{split}
        X^{E}(\mb p_0) = \frac{p_e \Pi +\beta \frac{S_0}{ N_0} }{\|{\textbf{X}(\mb p_0)}\|}, & \hspace{5mm}
          X^{I}(\mb p_0) = \frac{p_i \Pi  - (\tau_i + \mu + \delta) }{\|{\textbf{X}(\mb p_0)}\|},\\
           X^{A}(\mb p_0)= \frac{p_a \Pi}{\|{\textbf{X}(\mb p_0)}\|}, & \hspace{5mm}
           X^{U}(\mb p_0)= \frac{( p_e + p_i + p_a) \Pi + \beta \frac{S_0}{N_0}  - (\tau_i + \mu + \delta)}{\sqrt{3}\|{\textbf{X}(\mb p_0)}\|},
    \end{split}
\end{equation}
where $N_0 = S_0 + 1$ and
%
%\begin{equation*}
%\begin{split}
%    \norm{\textbf{X}(\mb p_0)}^2=& \left( p_s \Pi   - \left(\frac{\beta }{N_0} + v  + \mu\right) S_0\right)^2 + (p_v \Pi + v S_0)^2+
%     (p_r \Pi  + \tau_i)^2 \\ &+
%    \left( p_e \Pi + \frac{\beta }{N_0} S_0 \right)^2 +
%    (p_i \Pi  - (\tau_i + \mu + \delta))^2 +
%   (p_a \Pi )^2.
%\end{split}
%\end{equation*}
%
\begin{equation}\nonumber
 \begin{array}{rcl}
    \mathbf{X}(\mb p_0) &=&  
    \bigg( p_s \Pi   - \left(\frac{\beta }{N_0} + v  + \mu\right) S_0, \:  p_v \Pi + v S_0, \: p_r \Pi  + \tau_i, \vspace{2mm}\\
&& p_e \Pi +\beta \frac{S_0}{N_0}, \: p_i \Pi  - (\tau_i + \mu + \delta),\; p_a \Pi \bigg).
    \end{array}
\end{equation}

Taking $p_e, p_i, p_a > 0$ one gets 
$
    X^{I}(\mb p_0) < 0$ if and only if $p_i \Pi  - (\tau_i + \mu + \delta) < 0.
$
In the event of a first infected person in the symptomatic infected state $I$, this variable will decrease as long as the entry speed of symptomatic infected ($p_i \Pi$) is less than the net exit rate from this compartment ($\tau_i+ \mu+ \delta $).
Equivalently, the symptomatic infected population will decrease as long as the expected time of entry of an individual to the compartment $I$ is greater than the time of permanence in such state.
On the other hand, we have $       X^{E}(\mb p_0) > 0$ and $X^{A}(\mb p_0) > 0$. 
Hence, the arrival of a first infected person in the symptomatic state always induces an increase in the exposed and asymptomatic subpopulations.  
If we now consider the growth in a uniform direction to the three infectious states, from \eqref{Tasa_1_expuesto_SVEAIR} it is easy to see that:
\begin{equation*}
    X^{U}(\mb p_0) < 0  \Leftrightarrow
     \frac{(p_e + p_i + p_a) \Pi + \beta\frac{S_0}{N_0}}{(\tau_i + \mu + \delta)}    < 1.
\end{equation*}
As in the previous case, we can define an index analogous to the basic reproduction number $\mathcal{R}_{0}$ given by
\begin{equation}\label{R_I0}
  \mathcal{R}_{I_0}  = \frac{(p_e + p_i + p_a) \Pi + \beta\frac{S_0}{N_0}}{(\tau_i + \mu + \delta)}.
\end{equation}
Here, the subscript $I_0$ indicates that the first infected individual is in the symptomatic infectious state $I$.
Hence, $\mathcal{R}_{I_0}$ indicates the evolution of the disease at an initial instant upon the arrival of a first symptomatic infected person in a totally susceptible population and with immigration of infected people. If $ \mathcal{R}_{I_0} > 1$ then the disease is increasing, and if $ \mathcal{R}_{I_0} < 1$ the disease is decreasing.  
Thus, with a first symptomatic infected one has
\begin{equation*}
    X^{U}(\mb p_0) < 0 \Leftrightarrow \mathcal{R}_{I_0}<1.
\end{equation*}
Note that, if the proportions of entry of infected people ($p_e, p_i, p_a$), the total entry of people ($\Pi$) or the contact rate ($\beta$) increase, the value of $\mathcal{R}_{I_0}$ will grow. On the other hand, if the exit rates from the infected compartment are increased ---either due to recovery ($\tau_i$) or death ($\mu + \delta$)---, the value of $\mathcal{R}_{I_0 }$ decreases.
Therefore, the disease will be initially contained if the net rate of increase of individuals to the disease stages 
is less than the net rate of exit of infected from the system.

%%%%%%%%%%%%%%%%%%%%%%%%%%%%
\subsubsection{Patient zero in asymptomatic infected state $A$}
%%%%%%%%%%%%%%%%%%%%%%%%%%%%

Finally taking a first individual in the asymptomatic state $A$, that is,
% \begin{equation*}\label{primer_asintomático}
   $ \mb p_0 = (S_0, 0,0,0,0,1),$
%\end{equation*}
%
we obtain
\begin{equation}\label{Tasa_1_asintomatico_SVEAIR}
    \begin{split}
        X^{E}(\mb p_0) = \frac{p_e \Pi + \beta \xi \frac{S_0}{N_0}}{\|{\textbf{X}(\mb p_0)}\|}, &\hspace{5mm}
          X^{I}(\mb p_0) = \frac{ p_i \Pi + \tau_a (1-\alpha) }{\|{\textbf{X}(\mb p_0)}\|},\\
           X^{A}(\mb p_0)= \frac{ p_a \Pi   - (\tau_a + \mu + \delta)}{\|{\textbf{X}(\mb p_0)}\|}, &\hspace{5mm}
           X^{U}(\mb p_0)= \frac{( p_e + p_i + p_a) \Pi + \beta \xi \frac{S_0}{N_0} - (\alpha \tau_a + \mu + \delta)}{\|{\textbf{X}(\mb p_0)}\|},
               \end{split}
\end{equation}
where  $N_0 = S_0 + 1$ and 
\begin{equation}
 \begin{split}
    \textbf{X}(\mb p_0) =&  
    \bigg( p_s \Pi - \left(\frac{\beta \xi}{N_0} + v  + \mu\right) S_0,\:
     p_v \Pi + v S_0, \:
      p_r \Pi  + \alpha\tau_a ,\:
    p_e \Pi + \beta \xi \frac{S_0}{N_0},\\ \:
    & p_i \Pi + \tau_a (1-\alpha),\:
    p_a \Pi   - (\tau_a + \mu + \delta)\bigg).
    \end{split}
\end{equation}

Taking $p_e, p_i, p_a > 0 $ one obtains
$
             X^{A}(\mb p_0) <0$ if and only if $p_a \Pi   - (\tau_a + \mu + \delta) < 0.
$
That is, the rate of propagation in the direction of $A$ in the face of a first asymptomatic infected will be negative if and only if the rate of entry of asymptomatic individuals ($p_a \Pi$) is less than the exit rate from $A$ ($\tau_a + \mu + \delta$). We also have $
          X^{E}(\mb p_0) > 0,$ and
    $     X^{I}(\mb p_0) > 0$. 
This indicates that in the event of a patient zero in the asymptomatic infected state $A$ in a system with immigration of infected individuals, initially the exposed and symptomatic populations will increase. Now, if we consider the growth or decrease of the disease in a uniform direction, we have:
\begin{equation*}
        X^{U}(\mb p_0) <
        0 \Leftrightarrow  \frac{( p_e + p_i + p_a) \Pi + \beta \xi \frac{S_0}{N_0} }{(\alpha \tau_a + \mu + \delta)}  < 1.
\end{equation*}
Hence, we can define the threshold
\begin{equation}\label{R_A0}
  \mathcal{R}_{A_0} =   \frac{( p_e + p_i + p_a) \Pi + \beta \xi \frac{S_0}{N_0} }{(\alpha \tau_a + \mu + \delta)},
\end{equation}
where the index $A_0$ indicates that the first infected in the system is asymptomatic. %Si $ \mathcal{R}_{A_0} > 1$ la enfermedad en un instante inicial de tiempo se propaga y si  $ \mathcal{R}_{A_0} < 1$ decrece. El caso $ \mathcal{R}_{A_0} = 1$ indica que la enfermedad no crece ni decrece. 
 Note that  the increase in the proportions of infected people entering the system ($p_e, p_i, p_a$), the number of people entering the system ($\Pi$) or the contagion rate from asymptomatic to susceptible ($\beta \xi \kappa$) induce $\mathcal{R}_{A_0}$ to grow. On the other hand,  $\mathcal{R}_{A_0}$ decreases if the exit rates from the asymptomatic compartment increase, either due to recovery ($\alpha \tau_{a}$) or death ($\mu + \delta$).
 
 To complement these results on model \eqref{sistema_SVEAIR}, Appendix A presents an approach to define thresholds similar to \eqref{R_I0} and \eqref{R_A0} for general compartmental models.

%%%%%%%%%%%%%%%%%%%%%%%%%%%
\section{Case study: COVID-19 pandemic in Chile}
\label{sec:chile}
%%%%%%%%%%%%%%%%%%%%%%%%%%%

We will use model \eqref{sistema_SVEAIR} and the analytical results from the previous section to quantify the severity of the immigration of infected people into Chile.

%%%%%%%%%%%%%%%%%%%%%%%%%%%
\subsection{March 2020: Arrival of the first infected in Chile}
%%%%%%%%%%%%%%%%%%%%%%%%%%%

Let us consider the situation of Chile in March 2020, with a single first symptomatic infected person. The approximate population of Chile around that date was $19,200,000$~\cite{poblacion_chile_2020}. Therefore, we take the initial point as
\begin{equation}\label{punto_Chile_1inf}
    \mb p_0 = (19200000,0,0,0,1,0).
\end{equation}
Table \ref{Tabla_chile_1} shows the parameter values used in this scenario.
%the parameter values ​​used in this scenario.
%Todos los parámetros asociados  al virus SARS-Cov-2 y aquellos asociados a las proporciones de entrada a cada compartimento fueron tomados con los mismos valores de la tabla \ref{Tabla_USA_1i}. Mientras que, aquellos parámetros asociados exclusivamente al contexto de Chile en ese instante, que son la tasa de reclutamiento y la tasa de mortalidad natural, fueron obtenidos a partir de datos (ver referencias en la tabla).

\begin{table}[h]
%\centering
	\caption{Parameters values in March 2020 in Chile.}
	\label{Tabla_chile_1}
{\footnotesize
		%\begin{tabular}{ccccc}%%%The number of columns has to be defined here
		\begin{tabular*}{\textwidth}{@{\extracolsep\fill}lccccc}
			\toprule%
			\textbf{Parameter}   &     \textbf{Description} & \textbf{Value} & \textbf{Unit} & \textbf{Source}   \\
\midrule
$\Pi$ & Recruitment rate & 38000 &  day$^{-1}$ & \cite{nuevo_pudahuel,flight_trakking}\\
$v$ & Vaccination rate & 0 &  day$^{-1}$ &
assumed \\
$\omega$ &  Vaccination immunity loss rate & 0 & day$^{-1}$ & assumed\\
$\mu$ & Natural mortality rate & 6.35/1000 & day$^{-1}$ & \cite{mortalidad_chile}\\
$\beta$ & Contact rate & 0.09 &  day$^{-1}$ & \cite{Modelo_inmigracion}\\
$\sigma$ & Exposed exit rate & 0.13 &  day$^{-1}$ & \cite{Modelo_inmigracion}\\
$\alpha$ & Proportion of asymptomatic who recover & 0.14 &  day$^{-1}$ & \cite{Modelo_inmigracion}\\
$\phi$ & Proportion of exposed who become symptomatic &  0.7 & day$^{-1}$ & \cite{Modelo_inmigracion}\\
$\tau_a$ & Natural recovery rate of asymptomatic & 0.13978 & day$^{-1}$ & \cite{Modelo_inmigracion}\\
$\tau_i$ & Symptomatic recovery rate & 0.0833 & day$^{-1}$ & \cite{Modelo_inmigracion}\\
$\eta$ & Rate of loss of natural immunity &  0.011 & day$^{-1}$ & \cite{Modelo_inmigracion}\\
$\xi$ & Reduced transmission of asymptomatic &  0.3 & -- & \cite{Modelo_inmigracion}\\
$\delta$ & Disease mortality rate &  0.018/12 & day$^{-1}$ & \cite{Modelo_inmigracion}\\
$p_s$ & Entry rate of susceptible & 0.99 & percentage & assumed\\
$p_v$ & Entry rate of vaccinated & 0 & percentage & assumed\\
$p_r$ & Entry rate of recovered & 0 & percentage & assumed\\
$p_e$ & Entry rate of exposed & 0.006 & percentage & assumed \\
$p_i$ & Entry rate of symptomatic & 0.00268  & percentage & assumed \\
$p_a$ & Entry rate of asymptomatic & 0.00132 & percentage & assumed\\
\botrule
	\end{tabular*}}
\end{table}

Given that an entry flow of  infected people into the country is considered, the value of the standard $\mathcal{R}_0$ loses validity. Hence we use the formulas defined in \eqref{Tasa_1_infectado_SVEAIR} to measure the spread of the disease at that initial moment in time, thus obtaining:
$
      X^{E}(\mb p_0) =0.0027, \ 
        X^{I}(\mb p_0) =0.0012, \
        X^{A}(\mb p_0) = 0.0006.
$
This indicates that the disease in Chile had a growth trend in the three states associated with the disease. However, the increase was not yet of great magnitude because the rates are small, much closer to 0 than to 1.
Note that the exposed $E$ were growing $2.25$ times faster than the symptomatic infected $I$ and $4.5$ times faster than the asymptomatic $A$. 
On the other hand,  the rate of spread in a uniform direction to the three states associated with COVID-19 is
\begin{equation}
    \label{uniforme_1inf_chile}
    X^{U}(p_0) = 0.0026.
\end{equation}
And from \eqref{R_I0} we obtain
$ \mathcal{R}_{I_0} = 4170,$
which is clearly greater than 1. 

In conclusion, our indices and propagation rates in the model adapted to the Chilean reality in March 2020 coincide with what was observed at that time: COVID-19 in Chile was in the process of spreading, although in small magnitudes. However, small changes in the system parameters could have caused some of these rates to cross their relevant threshold values. In concrete, let us consider the scenario if preventive measures had been taken against COVID-19 in Chile before the appearance of this first symptomatic infected person.
Let us focus on the entry proportions associated with the exposed ($p_e$) and infectious ($p_i$) states in  \eqref{sistema1_tasa_unif}, along with the relationships
\begin{equation}\label{relaciones_papi_CHILE}
        p_a = \frac{0.33}{0.67} p_i, \hspace{3mm} 
        p_s + p_e + p_i + p_a = 1,\hspace{3mm}
        p_s, p_e, p_i, p_a \in [0,1].
\end{equation}

\begin{figure}[!htbp]
\centering
%\subfloat{\includegraphics[width=0.47\linewidth]{Imagenes/Tasaunif_1_pe_pi.png}\label{Chile_1f_beta}}
%\hspace{0.2cm}
%\subfloat{\includegraphics[width=0.5\textwidth]{Imagenes/Tasaunif_1_pe_pi_zoom.png}\label{Chile_1f_v}}
    \includegraphics[width=\textwidth]{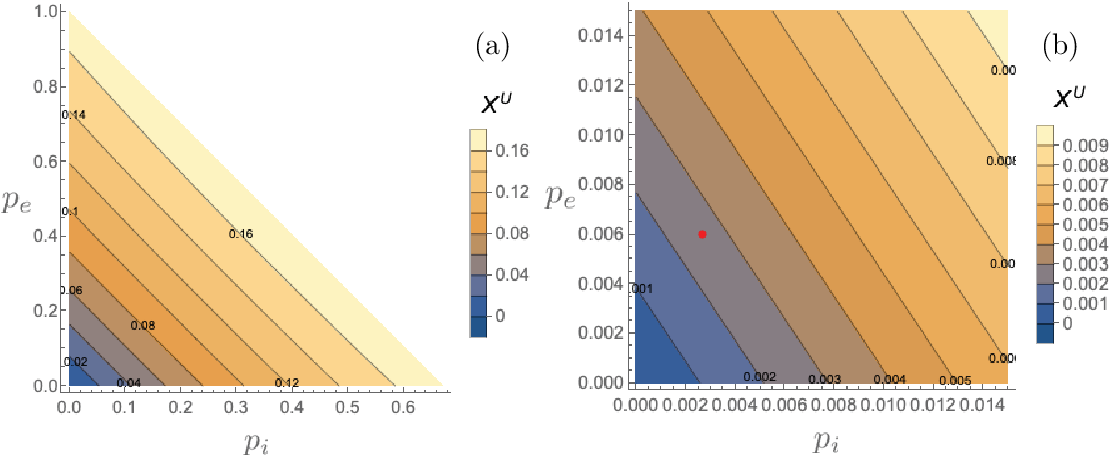}
\caption{Level curves of $X^{U}(p_0)$ defined in \eqref{sistema1_tasa_unif} evaluated at \eqref{punto_Chile_1inf} as a function of $p_e$ (a) and $p_i$ (b). In panel (a) the domain is defined by \eqref{relaciones_papi_CHILE}; and in (b) the domain reduces to $[0,0.015]\times [0,0.015]$. The rest of the parameters are fixed according to the values in table \ref{Tabla_chile_1}.}
\label{fig:Grafico_unif1_pe_pi_chile}
\end{figure}

Figure \ref{fig:Grafico_unif1_pe_pi_chile} shows level curves of the rate $X^{U}(p_0)$ \eqref{sistema1_tasa_unif} based on the entry proportions of exposed ($p_e$) and symptomatic infected ($p_i$).  On one hand, the lower the proportion of exposed and infected entries, the lower the speed of spread of the disease. At the beginning of the pandemic in Chile, the uniform propagation rate was between the level curves $0.002$ and $0.003$, specifically at the red dot in figure~\ref{fig:Grafico_unif1_pe_pi_chile}(b). Taking preventive measures or entry requirements at the Chilean border, such as a negative antibody detection test or avoiding the accumulation of passengers on the plane, could have slowed the spread.

\begin{figure}[htbp]
\centering
%\subfloat[]{\label{fig:Grafico_unif1_v_beta_Chile_a}\includegraphics[width=0.48\linewidth]{Imagenes/Tasaunif_1_beta.png}}\\
%\subfloat[]{\label{fig:Grafico_unif1_v_beta_Chile_b}\includegraphics[width=0.48\textwidth]{Imagenes/Tasaunif_1_v.png}}\hspace{0.3cm}
%\subfloat[]{\label{fig:Grafico_unif1_v_beta_Chile_c}\includegraphics[width=0.48\textwidth]{Imagenes/Tasaunif_1_v_zoom.png}}
    \includegraphics[width=\textwidth]{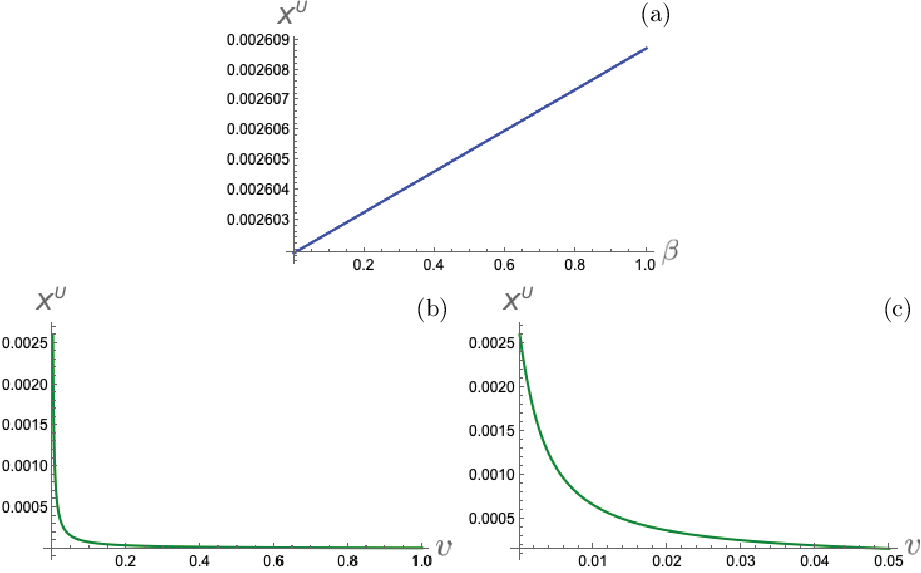}
\caption{The propagation rate $X^{U}(p_0)$ defined in \eqref{sistema1_tasa_unif} evaluated at \eqref{punto_Chile_1inf} as a function of: (a) the contact rate ($\beta$); (b) a (hypothetical) early vaccination rate ($v$) with an enlargement in panel (c) for $v\in[0,0.05]$. The rest of the parameters are fixed as in table \ref{Tabla_chile_1}.}
\label{fig:Grafico_unif1_v_beta_Chile}
\end{figure}

On the other hand, from figure~\ref{fig:Grafico_unif1_v_beta_Chile}(a) we see that the lower the contact rate $\beta$, the lower the propagation rate $X^U$. However, the impact that a quarantine or a capacity reduction could have had on the speed of spread of COVID-19 at that moment would not have been significant. Indeed, between $\beta=0$ and $\beta=1$ the rate of spread can vary at most $0.26\%$. 
Panels (b) and (c) of figure~\ref{fig:Grafico_unif1_v_beta_Chile} confirm that
adding a vaccination program available at the start of the pandemic would have helped contain the disease.  
In fact, from the figure we conclude that if the vaccination rate takes the value of $v=0.0036$ the spread rate decreases by half.

%%%%%%%%%%%%%%%%%%%%%%%%%%%%
\subsection{March 2022: Start of the vaccination program in Chile}
%%%%%%%%%%%%%%%%%%%%%%%%%%%%

We now take data from the 1st of March 2022 in Chile at the beginning of the national vaccination scheme.

Official numbers by the Chilean Ministry of Health show that $98,421$ confirmed cases were active on that date~\cite{casos_activos_Chile}.  
Since the proportion of confirmed cases of COVID-19 is divided into $67\%$ with symptoms and $33\%$ without symptoms~\cite{proporcion_asintomaticos}, we have that $65,943$ confirmed cases were symptomatic ($I$) and $32,478$ were asymptomatic ($A$).
In order to determine the number of people who had active vaccine immunity on March 1, 2022, we took the number of people who were vaccinated with the second dose or with a booster shot between December 1, 2021 and March 1, 2022, obtaining $5,402,398$ people in the vaccinated ($V$) compartment~ \cite{pasoapaso}. For this, we assumed that the vaccine has the potential to generate a 90-day immunity from the second dose as in  the $mRNA-1273$ immunizer developed by Moderna~ \cite{inmunidad_vacuna}. After this period of time, immunity is lost and the individual becomes susceptible again. 
For the calculation of the number of recovered individuals who have active natural immunity on March 1, 2022 in Chile, we considered that natural immunity lasts approximately 90 days. We took into account that  a patient is considered recovered after 12 days from the first symptom or positive PCR test and regarded the new confirmed cases from November 18, 2021 to February 16, 2022 from~\cite{pasoapaso}. 
Hence, a total of $950,000$ recovered ($R$) cases with active natural immunity were obtained. 
To determine the number of people exposed on March 1, 2022, we considered  all the new infections that were registered between March 2 and March 8, 2022, obtaining around 130,000 cases of estimated exposed ($E$) on March 1.
Since the approximate total population of Chile as of March 1, 2022 is $N=19,458,000$~\cite{poblacion_chile_2020}, the number of susceptible individuals is obtained from the formula $N = S + V +R+ E + I + A$, thus obtaining $S=13,826,231$.
In this way, initial conditions estimated for Chile on March 1, 2022 are:
\begin{equation}\label{punto_inicial_Chile}
\begin{array}{rcl}
    \mb p_0 &=& (S_0, V_0, R_0,  E_0, I_0, A_0) \\
    &=& (12.877.181, 5.402.398, 950.000, 130.000 , 65.943, 32.478).
    \end{array}
\end{equation}

\begin{table}[]
\centering
	\caption{Parameter values in March 2022 in Chile.}
	\label{Tabla2}
{\footnotesize
		\begin{tabular}{ccccc}%%%The number of columns has to be defined here
			\hline
			\textbf{Parameter}   &     \textbf{Description} & \textbf{Value} & \textbf{Unit} & \textbf{Source}   \\
\hline
$\Pi$ & Recruitment rate & 13.000 &  day$^{-1}$ &  \cite{nuevo_pudahuel,flight_trakking}  %
\\
$v$ & Vaccination rate & $66185/18275032$ &  day$^{-1}$&
\cite{pasoapaso,piramide_poblacion} 
\\
$\omega$ &  Vaccination immunity loss rate & 1/90 & day$^{-1}$ & \cite{inmunidad_vacuna,inmunidad_vacuna_2}
\\
$\mu$ & Natural mortality rate & 6.25/1000 & day$^{-1}$ & \cite{mortalidad_chile}\\
$\delta$ & Disease mortality rate & 124/100880 & day$^{-1}$&  \cite{pasoapaso}
\\
$p_s$ &  Entry rate of $S$ & 0.6 & percentage &  \cite{portal_transparencia_Chile} 
\\
$p_v$ & Entry rate of $V$ & 0.3 & percentage &  \cite{portal_transparencia_Chile}  
\\
$p_r$ & Entry rate of $R$ & 0.047077 & percentage &  \cite{portal_transparencia_Chile} 
\\
$p_e$ & Entry rate of $E$ & 0.0396923 & percentage &  \cite{portal_transparencia_Chile} 
\\
$p_i$ & Entry rate of $I$ & 0.0092615 & percentage &  \cite{portal_transparencia_Chile}  
\\
$p_a$ &Entry rate of $A$ & 0.0039692 & percentage & \cite{portal_transparencia_Chile}  
\\
$\beta$ & Contact rate & 0.09 &  day$^{-1}$ & \cite{Modelo_inmigracion} \\
$\sigma$ & Exit rate of exposed & 0.13 &  day$^{-1}$ & \cite{Modelo_inmigracion} \\
$\alpha$ & Proportion of $A$ who recover & 0.14 &  day$^{-1}$ & \cite{Modelo_inmigracion} \\
$\phi$ & Proportion of $E$ who become symptomatic &  0.7 & day$^{-1}$ & \cite{Modelo_inmigracion} \\
$\tau_a$ & Natural recovery rate of asymptomatic & 0.13978 & day$^{-1}$ & \cite{Modelo_inmigracion}  \\
$\tau_i$ & Natural recovery rate of symptomatic & 0.0833 & day$^{-1}$ & \cite{Modelo_inmigracion} \\
$\eta$ & Rate of loss of natural immunity  &  0.011 & day$^{-1}$ & \cite{Modelo_inmigracion} \\
$\xi$ & Reduced transmission of asymptomatic &  0.3 & -- & \cite{Modelo_inmigracion} \\
\hline
	\end{tabular}}
\end{table}

Table \ref{Tabla2} shows the parameter values of \eqref{sistema_SVEAIR} estimated as of March 1, 2022 in Chile. With this input the propagation rates \eqref{sistema1_tasa_expuestos}, \eqref{sistema1_tasa_infectados}, \eqref{sistema1_tasa_asintomática} and \eqref{sistema1_tasa_unif} take the following values:
\begin{equation}\label{tasas_2marzo2022}
 \begin{array}{cc}
     X^{E}(\mb p_0) = -0.178221,\ &
X^{I}(\mb p_0) = 0.138609,\\
X^{A}(\mb p_0) = 0.00476035,\ &
X^{U}(\mb p_0) = -0.0201213.
 \end{array}   
\end{equation}

%Some parameters of \eqref{sistema_SVEAIR} were estimated as of March 1, 2022 in Chile. : la tasa de reclutamiento o tasa de entrada al país ($\Pi$) - calculamos la cantidad de vuelos que llegaron desde el exterior el 1 de marzo del 2022 considerando además el modelo de avión, la tasa de vacunación ($v$) -  dividimos la cantidad de personas vacunadas el 1 de marzo del 2022 por la población objetivo de las vacunas en Chile, las proporciones de reclutamiento a cada compartimento $p_s, p_v, p_r p_e, p_i$ y $p_a$ - se estudiaron los requisitos de entrada en marzo de 2022 en Chile y se solicitó en el Portal Virtual de Transparencia del Estado de Chile \cite{portal_transparencia_Chile} la cantidad de personas que entraron al país con un PCR positivo mensualmente durante enero, febrero y marzo del 2022. Otros parámetros fueron obtenidos de estudios preexistentes como la tasa de pérdida de inmunidad de vacunación ($\omega$), la tasa de mortalidad natural ($\mu$)  y la tasa de mortalidad de la enfermedad ($\delta$). El resto de los parámetros fue obtenido de la tabla \ref{Tabla_USA_1i}.  Luego, para este análisis se utilizaron los valores de la Tabla \ref{Tabla2}. 

%Dado que se considera la entrada de infectados al sistema ($p_e, p_a, p_i > 0$), el indicador $\mathcal{R}_0$ pierde validez como umbral de crecimiento o decrecimiento. Por lo tanto, con los valores de parámetros definidos en la tabla \ref{Tabla2} y el punto inicial \eqref{punto_inicial_Chile}, se calcularon 

According to \eqref{tasas_2marzo2022}, the disease on March 1, 2022 was growing in the direction of both symptomatic $I$ and asymptomatic infected $A$. In fact, the symptomatic population was growing 29 times faster than the asymptomatic. On the other hand, the spread of COVID-19 at that time was decreasing in the direction of the exposed $E$. And, if we put these three states together, the value of $X^U<0$ indicates that the whole cluster of disease variables was decreasing.

Let us now investigate the effect of changing the entry proportions on  the propagation rates \eqref{sistema1_tasa_expuestos}, \eqref{sistema1_tasa_infectados}, \eqref{sistema1_tasa_asintomática} and \eqref{sistema1_tasa_unif}. For this we consider $ p_s ,p_e , p_i , p_a \in [0,1]$ and the restrictions~\cite{proporcion_asintomaticos,proporcion_expuestos_infectados}:
\begin{equation}\label{proporción_relaciones_pepapi}
    p_e = 2.5 p_i,  \hspace{3mm}
   p_a = \frac{0.33}{0.67}p_i,\hspace{3mm}
    p_s = 1- (p_v + p_r + p_e + p_i + p_a).
\end{equation}
In this way, by increasing, for example, the entry rate of symptomatic infected $p_i$, the entry ratios of exposed $p_e$ and asymptomatic $p_a$ must also increase and the entry proportion of susceptible $p_s$ will decrease according to \eqref{proporción_relaciones_pepapi}.  This allows the following analysis below to be more realistic.

 \begin{figure}[htbp]
\centering
%\subfloat[]{\label{Chile_inf_a}\includegraphics[scale=0.5]{Imagenes/Tasa_inf_pi-Pi.png}}\hspace{2mm}
%\subfloat[]{\label{Chile_inf_b}\includegraphics[scale=0.5]{Imagenes/Tasa_inf_pi-Pi_zoom.png}}\\
%\subfloat[]{\label{Chile_inf_c}\includegraphics[scale=0.5]{Imagenes/Tasa_inf_beta-v.png}}\hspace{2mm}
%\subfloat[]{\label{Chile_inf_d}\includegraphics[scale=0.5]{Imagenes/Tasa_inf_beta-v_zoom.png}}
    \includegraphics[width=\textwidth]{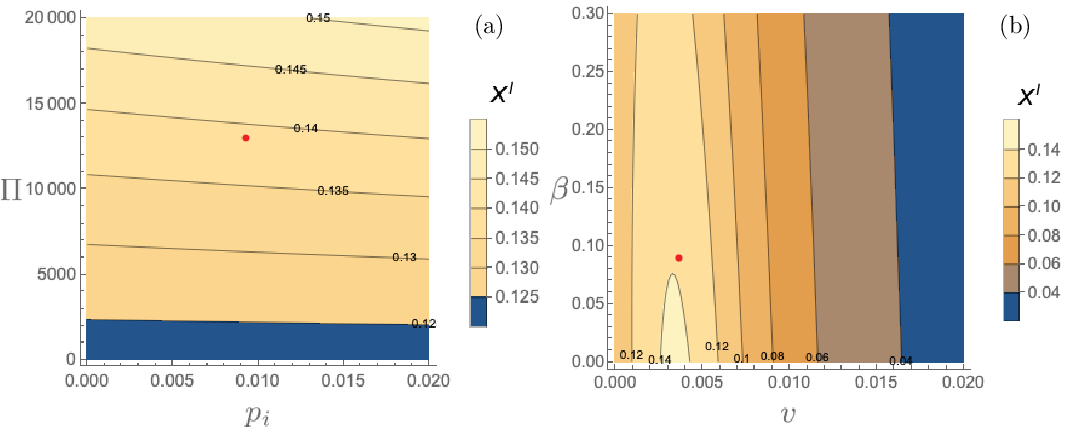}
\caption{Level sets of $X^{I}(p_0)$ defined in \eqref{sistema1_tasa_infectados} evaluated at \eqref{punto_inicial_Chile}: (a)  As a function of the proportion of recruitment to infected ($p_i$) and recruitment rate to the country ($\Pi$); (b) As a function of the vaccination rate ($v$) and contact rate ($\beta$). The red dot indicates the value of the parameters and value of the rate $X^I (\mb p_0)$ at the instant of evaluation~\eqref{punto_inicial_Chile}. The other parameters are fixed according to the values in table \ref{Tabla2}.}
\label{fig:CP_Chile_inf}
\end{figure}

Figure \ref{fig:CP_Chile_inf} shows the level sets of $X^{I}(\mb p_0)$ with respect to $(p_i,\Pi)$ in panel (a), and $(v,\beta)$ in panel (b). 
In both images the red dot represents the state of the system at the moment of evaluation \eqref{punto_inicial_Chile}. It follows from figure~\ref{fig:CP_Chile_inf}(a) that regardless of the number of people entering the country ($\Pi$) and their proportion of infected ($p_i$), the number of symptomatic infected $I$ will grow since $X^{I}(\mb p_0)$ remains positive. However, the lower $\Pi$  and the lower $p_i$, the lower the magnitude of their spread. Note that the propagation could have been reduced by partially closing the borders, thus reducing the number of people entering the country ($\Pi$). If the entry of people into the country had been reduced by half, $X^{I}(\mb p_0)$ would have decreased by approximately $6.22\%$ and, if the borders were permanently closed, $X^{I}(\mb p_0)$ would have decreased by $11.8\%$. On the other hand, from figure~\ref{fig:CP_Chile_inf}(b) it can be observed that that the number of infected people will increase as long as the entry rate of these individuals into the system is maintained, although the speed of increase remains relatively small. Also, one may conclude that increasing the vaccination rate $v$ is a much more effective measure to reduce the speed of spread of those infected at that moment. Indeed, if the vaccination rate increases to $v=0.008$ (i.e, to vaccinate 8 people per 1,000 per day),  $X^{I}(\mb p_0)$ decreases by $36.5\%$.

Let us now see what is the impact on the rate $X^{U}(\mb p_0)$ of varying some parameters of interest. From~\eqref{tasas_2marzo2022}, the rate in the uniform direction $X^{U}(\mb p_0)$ on March 1, 2022 in Chile was close to $-0.02$. In other words, grouping the three states associated with the disease, COVID-19 at that time was decreasing. However, the magnitude of this rate is very close to 0, so we would be interested in increasing this rate of decline even more, and analyzing under what conditions it could become positive.

%%%%%%%%%%%%%%%%%%%%%%%%%%%%
\subsubsection{Border measures}
 %%%%%%%%%%%%%%%%%%%%%%%%%%%%

 \begin{figure}[h!]
\centering
%\subfloat[]{\label{Chile_unif_frontera_a}\includegraphics[scale=0.49]{Imagenes/Tasa_unif_pi-Pi_zoom.png}} \hspace{1cm}
%\subfloat[]{\label{Chile_unif_frontera_b}\includegraphics[scale=0.49]{Imagenes/Tasa_unif_pi-pa.png} }%\hspace{2mm}
%\subfloat[]{\label{Chile_unif_frontera_c}\includegraphics[scale=0.49]{Imagenes/Tasa_unif_pi-pa_zoom.png} }
 %\\
%\subfloat[]{\label{Chile_unif_frontera_d}\includegraphics[scale=0.49]{Imagenes/Tasa_unif_pi_pv.png} }
%\subfloat[]{\label{Chile_unif_frontera_e}\includegraphics[scale=0.49]{Imagenes/Tasa_unif_pi_pv_zoom.png} }
    \includegraphics[width=\textwidth]{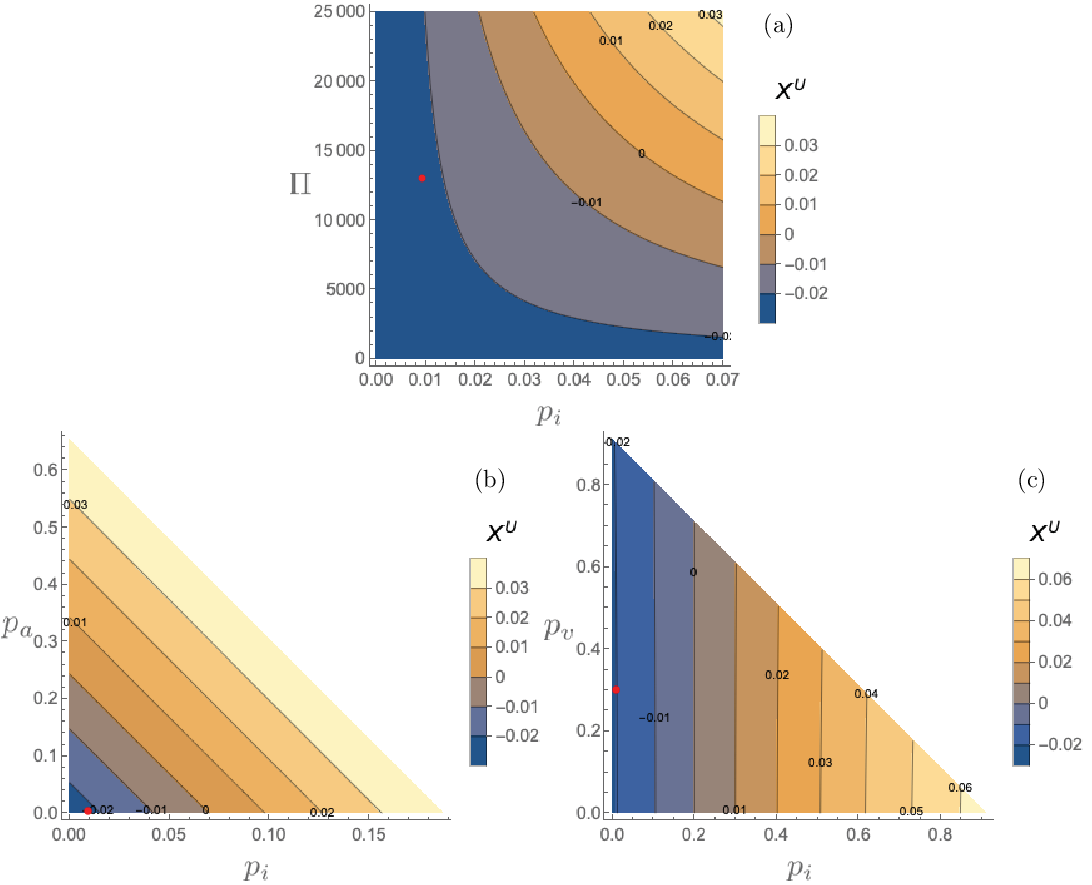}
\caption{Contour plots of $X^{U}(\mb p_0)$ defined in \eqref{sistema1_tasa_unif} evaluated at \eqref{punto_inicial_Chile} as a function of parameters related to border measures: (a) As a function of the ratio of recruitment to infected ($p_i$) and recruitment rate to the whole country ($\Pi$); (b) As a function of the entry rate of symptomatic infected ($p_i$) and the entry rate of asymptomatic ($p_a$); (c) As a function of $p_i$ and the entry rate of vaccinated ($p_v$). The red dot indicates the value of parameters and value of $X^U (\mb p_0)$ at the moment of evaluation~\eqref{punto_inicial_Chile}. The other parameters are fixed according to the values in table \ref{Tabla2}.}
\label{Chile_frontera}
\end{figure}

Figure \ref{Chile_frontera} shows the uniform propagation rate \eqref{sistema1_tasa_unif} varying some parameters associated with possible measures that could have been taken on the Chilean border. The entry of people to Chile on March~1, 2022 was around $13,000$ per day and the entry rate of infected was less than $0.01$ approximately, see red dot in figure~\ref{Chile_frontera}(a). Although this point is in a zone of parameters that allows the disease to decrease, reducing the entry rate of infected $p_i$ and/or the number of people entering the country $\Pi$ would ensure that the disease is kept decreasing and at a higher speed. However, at that time the measures taken at the Chilean borders were becoming less strict.
In the short term, more people were beginning to enter the country and a negative PCR test was no longer a requirement to enter, which translates into an increase in the parameters $\Pi$ and $p_i$; eventually, this would cause an increase in $X^{U}(\mb p_0)$, see again figure~\ref{Chile_frontera}(a). Hence, in hindsight, the growth of the speed of spread of the disease at that instant would have been predicted by the index $X^U(\mb p_0)$ as shown in figure~\ref{Chile_frontera}(a). The other panels in Figures~\ref{Chile_frontera} take into account the relations \eqref{proporción_relaciones_pepapi} to plot $X^{U}(\mb p_0)$ as a function of some of the entry rates into the country ($p_i, p_a, p_v$).
It follows that although decreasing the proportions of infected and asymptomatic immigrants simultaneously causes a decrease in the propagation rate, to reduce the entry rate of symptomatic infected (rather than asymptomatic ones) has a greater impact; see figure~\ref{Chile_frontera}(b). A similar result is obtained from the figure~\ref{Chile_frontera}(c); namely, it is more convenient to keep a check on the proportion of infected inmigrants than on the vaccinated ones. In fact, if the entry rate of infected $p_i$ increases above $0.2$, $X^{U}(\mb p_0)$ becomes positive independent of the entry rate of vaccinated $p_v$. In other words, if we consider all the other parameters fixed, if more than $2\%$ of the people entering the country are  symptomatic infected ($I$) of COVID-19, the disease within the country will increase regardless of the number of vaccinated immigrants. Therefore, given this observation, a negative PCR test at the entrance of the country emerges as a more effective measure at the Chilean border than a vaccination certificate.

%%%%%%%%%%%%%%%%%%%%%%%%%%%%
\subsubsection{Contagion preventive measures}
%%%%%%%%%%%%%%%%%%%%%%%%%%%%

\begin{figure}[h]
\centering
%\subfloat[]{\label{Chile_unif_preventivas_a}\includegraphics[scale=0.5]{Imagenes/Tasa_unif_beta-v.png}}\hspace{2mm}
%\subfloat[]{\label{Chile_unif_preventivas_b}\includegraphics[scale=0.5]{Imagenes/Tasa_unif_beta-v_zoom.png}}
    \includegraphics[scale=0.7]{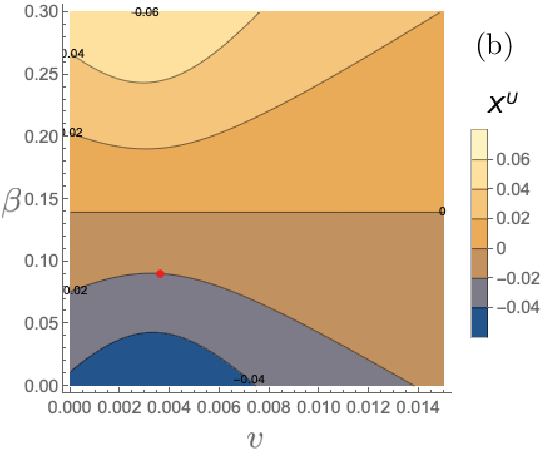}
\caption{ (a) Contour plots of $X^{U}(\mb p_0)$ defined in \eqref{sistema1_tasa_unif} evaluated at \eqref{punto_inicial_Chile}  as a function of the vaccination rate ($v$) and contact rate ($\beta$). The red dot indicates the value of parameters and value of $X^U (\mb p_0)$ at the moment of evaluation~\eqref{punto_inicial_Chile}. The other parameters are fixed according to the values in table \ref{Tabla2}.}
\label{fig:Chile_unif_preventivas}
\end{figure}

Figure \ref{fig:Chile_unif_preventivas} aims to capture the impact of taking hypothetical preventive measures of contagion on the uniform propagation rate $X^{U}(\mb p_0)$ on March 1, 2022. According to the model, there are two parameters associated with contagion prevention: the contact rate ($\beta$) and the vaccination rate ($v$). From figure~\ref{fig:Chile_unif_preventivas} we see that there is a threshold for the contact rate around $\beta \approx 0.1395$. That is, if the contact rate exceeds this value, $X^{U}(\mb p_0)$ becomes positive, independent of the vaccination rate. Therefore, at that moment the disease could go from decreasing to increasing if the contact between susceptible people with infected people (with or without symptoms) has a sharp increase in a short period of time. However, if $\beta$ exceeds the threshold value, it is preferable that there be a higher vaccination rate, since the growth rate of the disease is lower, see the enlargement in figure~\ref{fig:Chile_unif_preventivas}. Additionally, we can see from the red dot in figure~\ref{fig:Chile_unif_preventivas} that, if we keep the contact rate constant and increase only the vaccination rate, $X^{U}(\mb p_0)$ increases. This seems counterintuitive at first but it is not contradictory since $X^{U}(\mb p_0)$ continues to be negative, that is, the disease continues to decrease, only at a slightly slower speed since fewer people will become sick.

%%%%%%%%%%%%%%%%%%%%%%%%%%%%
\subsubsection{Measures associated with strain parameters}
 %%%%%%%%%%%%%%%%%%%%%%%%%%%%

\begin{figure}[h]
\centering
%\subfloat[]{\label{Chile_cepa_a}\includegraphics[scale=0.5]{Imagenes/Tasa_unif_delta-tau.png}}\hspace{2mm}
%\subfloat[]{\label{Chile_cepa_b}\includegraphics[scale=0.5]{Imagenes/Tasa_unif_delta-tau_zoom.png}}
    \includegraphics[width=\textwidth]{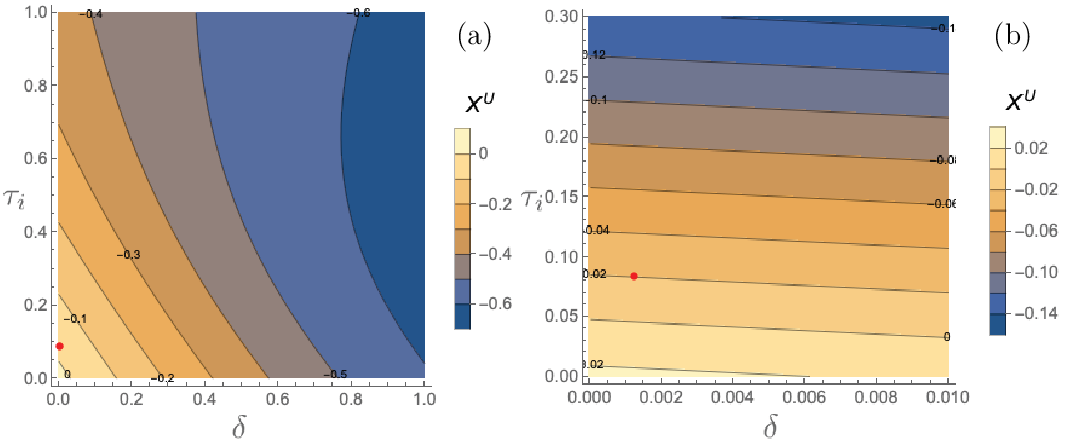}
\caption{(a) Level sets of $X^{I}(p_0)$ defined in \eqref{sistema1_tasa_infectados} evaluated at \eqref{punto_inicial_Chile} based on parameters related to the parameters of the COVID-19 strain, namely, the mortality rate of the disease ($\delta$) and recovery rate of infected ($\tau_i$); (b) The same plot as (a) in a smaller region of parameters.  The red dot indicates the value of parameters and value of $X^U (\mb p_0)$ at the moment of evaluation~\eqref{punto_inicial_Chile}. The other parameters are fixed according to the values in table \ref{Tabla2}.}
\label{fig:Chile_unif_cepa}
\end{figure}

Finally, in figure~\ref{fig:Chile_unif_cepa} two model parameters associated with the COVID-19 strain were taken as variables: the mortality rate associated with the disease ($\delta$) and the recovery rate of symptomatic infected ($\tau_{i}$).
In figure~\ref{fig:Chile_unif_cepa}(a) we obtain that the higher $delta $ and $\tau_i$, the lower the value of the uniform propagation rate.This is because the model \eqref{sistema_SVEAIR} considers that the higher the mortality rate of the disease, the fewer individuals are in a state associated with the disease and, the higher $\tau_{i}$, the fewer days an individual remains on average in a symptomatic infectious state. In this case study, the mortality rate was less than $0.0013$ and the recovery rate was increasing. The number of days in which a person in Chile was considered to be in a symptomatic infectious state was changing from 14 days to 7 days, which translates into a value of $\tau_i = 0.14$ approximately. It follows from the enlargement in figure~\ref{fig:Chile_unif_cepa}(b)  given this increase in $\tau_i$, $X^{U}(\mb p_0)$ increases its rate of decrease, remaining between the level curves $-0.04$ and $-0.06$. In other words, the rate of decrease of the disease increases by more than $50\%$ if the strain has a recovery time of 7 days less than the initial strain.

%%%%%%%%%%%%%%%%%%%%%%%%%%%%%%%%%%%%%%%
\section{Asymptotic properties of the propagation rate}
\label{sec:asymptotic}
%%%%%%%%%%%%%%%%%%%%%%%%%%%%%%%%%%%%%%%

In this section we present results on the asymptotic behavior of the propagation rates \eqref{eq:v} and \eqref{eq:promediot} in a general epidemiological system with $d=n+m$ compartments like the one defined in (\ref{eq:general}).

Let $\mb p^*$ be a hyperbolic attracting equilibrium point of (\ref{eq:general}), i.e., all the orbits in a sufficiently small neighborhood remain close to $\mb p^*$ for $t>0$ and converge asymptotically to $\mb{p}^*$ as $t\rightarrow\infty$~\cite{GH}.
Let $A$ be the Jacobian matrix of (\ref{eq:general})  evaluated at $\mb p^*$. 
Without loss of generality, we can assume the equilibrium point $\mb p^*$ to be at the origin $\mb 0 \in \mathbb{R}^{d}$ and rewrite (\ref{eq:general}) locally in a sufficiently small  neighborhood $\mathcal{V}$ of $\mb p^*$ in the form
     \begin{equation}\label{sistema_n+m_origen}
        \mb X:  \begin{cases}
          \begin{pmatrix}
          \mb x'\\
          \mb y'
          \end{pmatrix} = A \begin{pmatrix}
          \mb x\\
          \mb y
          \end{pmatrix} + H(\mb x,\mb y),
       \end{cases}
     \end{equation}
where $H: \mathbb{R}^{d} \to  \mathbb{R}^{d} $ is $C^r$, $r\geq1$, and satisfies $H(\mb p^*)=\mb 0$ and $DH(\mb p^*)=0$ in $\mathcal{V}$.

We will show that the asymptotic behavior of \eqref{eq:v} and  \eqref{eq:promediot} depends on the nature of the leading eigenvalues of $A$ in \eqref{sistema_n+m_origen}. Our analysis will be restricted to the generic case in which there is one (simple) real leading eigenvalue.
Hence, we can order the eigenvalues of $A$ in the following way:
 $${\rm Re}(\lambda_{d}) \leq {\rm Re}(\lambda_{d-1}) \leq ...\leq {\rm Re}(\lambda_2) < {\rm Re}(\lambda_{1}) < 0,$$
 where  $\lambda_1\in\mathbb{R}$ is  the leading eigenvalue of $A$.
Let $\mb v_{i} \in \mathbb{R}^{d}$, $i=1,...,d$, be the associated (generalized) eigenvectors. 
 Without loss of generality, let us assume that $\mb v_1$ is a unit eigenvector of $\lambda_1$, i.e., $\|\mb v_1\|=1$.

%%%%%%%%%%%%%%%%%%%%%%%%%%%%%%
% \subsection{Case 1: One real leading eigenvalue}
% \label{sec:casoreal}
 %%%%%%%%%%%%%%%%%%%%%%%%%%%%%%

If $\mathcal{B}(\mb p^*)$ is the basin of attraction of $\mb{p}^*$, 
 let us denote as $\mathcal{B}_{\rm loc}=\mathcal{B}(\mb p^*)\cap \mc V$ to the ``local'' basin 
 in the neighbourhood $\mc V$.
 Since $\mb p^*$ is hyperbolic, almost every orbit in $\mathcal{B}_{\rm loc}$ converge to $\mb p^*$ tangent to the leading stable space $\mathcal{E}^{l} = \langle \mb v_1 \rangle$. The exception are those orbits lying in the local (non-leading)  strong stable manifold $W^{ss}_{\rm loc}(\mb p^*)\subset \mathcal{B}(\mb p^*)$ of dimension $d-1$ which converge tangent to the strong stable eigenspace $\mathcal{E}^{ss} = \langle \mb v_2,...,\mb v_d \rangle$. 
 % On the other hand, the orbits of those points in the basin of attraction of $\mb p^*$ that are outside $W_{\rm loc}^{ss}(\mb p^*)$ converge tangent to the weak stable space $\mathcal{E}^{l} = \langle \mb v_1 \rangle$. 
  Moreover, $\mathbb{R}^{d} = \mathcal{E}^{ss} \oplus \mathcal{E}^{l}$; see~\cite{GH,Teoria_cualitativa_sistemas} for further details. In what follows, to ease the notation, we drop the `$\rm loc$' symbol in both $\mathcal{B}_{\rm loc}$ and $W_{\rm loc}^{ss}(\mb p^*)$.

    Let $\mb p_{0} \in \mathcal{B}\setminus W^{ss}(\mb p^*)$ be a point in the basin of $\mb p^*$ lying outside the strong stable manifold.
   Since the solution $\varphi_{\mb p_0}(t)$ of \eqref{sistema_n+m_origen} remains in $\mathcal{V}$ for all $t>0$,  we can write the tangent vector field to the integral curve $\varphi_{\mb p_0}(t)$ in $\mathcal{V}$ in the form:    
\begin{equation}\label{eq:v_tangente}
 \textbf{X}(\varphi_{\mb p_0}(t)) = \frac{d\varphi_{\mb p_0}(t)}{dt} = 
\sum_{j=1}^{d} \sigma_{j}(t)\mb v_j,
\end{equation}
for certain coefficients $\sigma_j(t),$ $j=1,\ldots,d.$ More specifically, if $\lambda_j\in\mathbb{R}$ is a real eigenvalue with multiplicity one, we have
$$\sigma_j (t) = C_j \lambda_j e^{\lambda_j t} + h_{j}(t). $$
Here $C_j \lambda_j e^{\lambda_j t}$ corresponds to the component associated to the linear part of $\textbf{X}(\varphi_{\mb p_0}(t))$ along $\mb v_j$
and $h_{j} = K_j \langle H, \mb v_{j} \rangle$  is the nonlinear part along $\mb v_j$, where $C_j$ and $K_j$ are constants specified by the initial condition $\mb p_0$. 
 On the other hand, if $\lambda_{j,j+1} = \alpha \pm \beta i$, $j\neq1$, are (simple) complex conjugate eigenvalues, then
\begin{equation*}
  \sigma_{l}(t) =  \omega_{l}(t) + h_l(t), \:\:\:{\rm for}\,\,\, l=j,j+1, \\
\end{equation*}
with 
\begin{equation}
    \begin{split}
        \omega_{j}(t) &= e^{\alpha t} (\alpha C_j \cos(\beta t) - \beta C_j \sin(\beta t) + \alpha C_{j+1} \sin(\beta t) + C_{j+1} \beta \cos(\beta t)), \\
        \omega_{j+1}(t) &= e^{\alpha t} (-\alpha C_j \sin(\beta t) - \beta C_j \cos(\beta t) + \alpha C_{j+1} \cos(\beta t) - C_{j+1} \beta \sin(\beta t)),
    \end{split}
\end{equation}
where again $h_{l} = K_l \langle H, \mb v_{l} \rangle$ with $l= j,j+1$, and $C_l,K_l$ depend on the initial condition.

%where  $$\sigma_1(t)=  C_1 \lambda_1 e^{\lambda_1 t} + h_1(t).$$ where $C_1 \lambda_1 e^{\lambda_1 t}$ corresponds to the component associated to the linear part of $\textbf{X}(\varphi_{\mb p_0}(t))$ along $\mb v_1$ and  $h_{1} = K_1 \langle H, \mb v_{1} \rangle$ is the nonlinear part along $\mb v_1$; here $C_1$ and $K_1$ are constants associated with the initial condition $\mb p_0$.  Moreover, 

%In order to unify notation in \eqref{eq:v_tangente}, let us denote
%
% $$\sigma_1(t)=  C_1 \lambda_1 e^{\lambda_1 t} + h_1(t).$$
 %
 
Since the orbit $\varphi_{\mb p_0}(t)$ lies outside of $W^{ss}(\mb p^*)$, the tangent vector $\textbf{X}(\varphi_{\mb p_0}(t))$ in \eqref{eq:v_tangente} always has a nonzero component in the direction of $\mb v_1$, that is, $\sigma_{1}(t)\neq 0$ for all $t>0$.

Substituting \eqref{eq:v_tangente} in \eqref{eq:v}  we have
\begin{equation}\label{tasa_temporal_formula_3D}
\begin{split}
  \lim_{t \to \infty} X^{\textbf{u}}_{\mb p_0}(t) &= 
  \lim_{t \to \infty} \frac{\langle \textbf{X}(\varphi_{\mb  p_0}(t)), \textbf{u} \rangle}{\|{\textbf{X}(\varphi_{\mb p_0}(t))}\|}\\
 & = 
  \lim_{t \to \infty} \frac{\langle \sigma_1(t)  \mb v_1 +  \sum_{i=2}^{d} \sigma_{i}(t)\mb v_i, \textbf{u} \rangle}{\|{ \sigma_1(t)  \mb v_1 +  \sum_{i=2}^{d} \sigma_{i}(t)\mb v_i}\|}\\
  &=  \lim_{t \to \infty} \frac{\sigma_{1}(t)}{|\sigma_{1}(t)|}\frac{\langle   \mb v_1 +   \frac{ 1}{\sigma_1(t)}  \sum_{i=2}^{d} \sigma_{i}(t)\mb v_i  , \textbf{u} \rangle}{\|{ \mb v_1 +  \frac{ 1}{\sigma_1(t)}  \sum_{i=2}^{d} \sigma_{i}(t)\mb v_i  }\|}\\
%  \left\| \mb v_1 +   \frac{ 1}{\sigma_1(t)}  \sum_{i=2}^{d} \sigma_{i}(t)\mb v_i  \right\|^{-1} \langle   \mb v_1 +  \frac{ 1}{\sigma_1(t)}  \sum_{i=2}^{d} \sigma_{i}(t)\mb v_i , \textbf{u} \rangle\\
   &=  \lim_{t \to \infty} {\rm sign}\left(\sigma_{1}(t)\right)\frac{\langle   \mb v_1 +   \frac{ 1}{\sigma_1(t)}  \sum_{i=2}^{d} \sigma_{i}(t)\mb v_i  , \textbf{u} \rangle}{\|{ \mb v_1 +  \frac{ 1}{\sigma_1(t)}  \sum_{i=2}^{d} \sigma_{i}(t)\mb v_i  }\|}.
\end{split}
\end{equation}
By definition, the orbit $\varphi_{\mb p_0}(t)$ approaches $\mb p^*$ tangent to $\mathcal{E}^{l}$. It follows that 
\begin{equation}\label{ec:3D_caso1_argumento_tangente}
     \lim_{t \to \infty} \frac{\sigma_{k}(t)}{\sigma_{1}(t)} = 0, \ \ k=2,\ldots,d,
\end{equation}
for all the components $\sigma_{k}(t)$ of  the tangent vector $\textbf{X}(\varphi_{\mb p_0}(t))$ in \eqref{eq:v_tangente} associated with  $\mb v_k$, $k=2,\ldots,d$. %Note that, otherwise,  the orbit $\varphi_{\mb p_0}(t)$ would have 
%
%Indeed, the components of  $\textbf{X}(\varphi_{\mb p_0}(t))$ in \eqref{eq:v_tangente} associated with  $\mb v_k$, $k=2,\ldots,d$, converge to the origin faster than the component associated with $\mb v_1$. That is:
 %
%
Additionally, the sign of $\sigma_1(t)$ defines the direction of convergence of the orbit $\varphi_{\mb p_0}(t)$ to the equilibrium point: either along $+\mb v_1$ or  $-\mb v_1$. Moreover, shrinking the neighbourhood $\mc V$ if necessary, the sign of $\sigma_1(t)$ is completely determined by its value at $t=0$. 
 Hence we define:
\begin{equation}\label{eq:s}
s =  {\rm sign}(\sigma_1(t))  = {\rm sign}(\sigma(0)) = {\rm sign}(C_1 \lambda_1 + h_1(0) ).  
\end{equation}

Returning to (\ref{tasa_temporal_formula_3D}) we obtain the following result.

\begin{lemma}\label{lemma6.1} Let $\mb p^*$ be a hyperbolic attracting equilibrium of the system (\ref{sistema_general_1}). Let ${\rm Re}(\lambda_{d}) \leq {\rm Re}(\lambda_{d-1}) \leq ... \leq {\rm Re}(\lambda_2)< \lambda_{1} < 0 $ be the eigenvalues of $\mb p^*$.  Suppose that the eigenvector associated with $\lambda_1$ satisfies $\|{{\bf v}_1}\| = 1$. Then, there exists a sufficiently small neighborhood $\mc V$ of $\mb p^*$ such that
\begin{equation}\label{ec:2D_real_resultado_promedio_temporal_pre}
      \lim_{t\to \infty} X^{\mb{u}}_{\mb p_0}(t) =  s \langle  \mb v_1, {\bf u} \rangle,
\end{equation}
for all  $\mb p_{0} \in \mathcal{B}\setminus W^{ss}(\mb p^*)$ lying outside the strong stable manifold in the local basin $\mathcal{B}=\mathcal{B}(\mb p^*)\cap \mc V$ of $\mb p^*$,
%$\mb p_0\in \mathcal{B}(\mb p^*)\cap \mc V\setminus W^{ss}(\mb p^*)$, 
and where $s$ is given by \eqref{eq:s}.
\end{lemma}

The result (\ref{ec:2D_real_resultado_promedio_temporal_pre}) tells us that the propagation rate associated with generic solutions of (\ref{sistema_general_1}) in $\mathcal{B}\setminus W^{ss}(\mb p^*)$ for $t\gg1$ depends on three ingredients: 
\begin{itemize}
    \item The leading eigenvector associated with the attractor equilibrium point ($\mb v_1$).
    \item The chosen unit vector ($\textbf{u}$).
    \item The location of the initial condition $\mb p_0$ in $\mathcal{B}$, specified by $s$ in \eqref{eq:s}.% = {\rm sign}(C_1\lambda_1 + h_1(0))$.
\end{itemize}

If we consider the unit normal vector in the general form  \eqref{eq:u-general} and denote the leading eigenvector as
\begin{equation*}
        \mb v_{1} = \left(% \begin{matrix}
a_1,
\ldots,
a_n,
b_1,
\ldots,
b_m
%\end{matrix}
 \right)^t,
    \end{equation*}
     with $\|{\mb v_1}\| = 1 $, (\ref{ec:2D_real_resultado_promedio_temporal_pre}) turns into
    \begin{equation}\label{ec_tasa_tiempo_n+m}
         \lim_{t \to \infty} X^{\textbf{u}}_{\mb p_0}(t) =  s (u_1b_1 + ... + u_m b_m).
    \end{equation}
    
If $b_i=0$ for every $i=1,\ldots,m$, then the (asymptotic) orbit is parallel to $\mc S$ and, hence, $\lim_{t \to \infty} X^{\textbf{u}}_{\mb p_0}(t) =0.$ Let us assume that at least some $b_i\neq0$; without loss of generality, suppose that $b_m\neq0$.
In particular, let us choose $+\mb v_1$ such that $b_m>0$; i.e., the component of $+\mb v_1$ in the direction of the state variable $y_m$ is positive.
Notice that the value of $s$ in \eqref{eq:s} and \eqref{ec_tasa_tiempo_n+m} is determined by the direction in which the orbit converges to $\mb p^*$, either along $+\mb v_1$ or $-\mb v_1$. Hence, $s$ is specified by the component of the tangent vector to the orbit in the direction of $y_m$, i.e., $s={\rm sign}(\sigma_1(0))={\rm sign}(y_{m}'(\mb p_0)) $. Moreover, if $\mc{V}$ is sufficiently small, the sign of $y_{m}'(\varphi_{\mb p_0}(t))$ remains the same for all $t\geq0$.

\begin{definition}\label{def:cuencas_n+mD}
Given system (\ref{sistema_general_1}), let $\mc B$ be as in Lemma~\ref{lemma6.1}. For $\tau\gg1$ we define the halfbasins $\mathcal{B}_{+}$ and $\mathcal{B}_{-}$ of $\mb p^*$ as
$$\mathcal{B}_{+} = \{ \mb p \in \mathcal{B}\setminus W^{ss}(\mb p^*)\, |  \,\,\, y_{m}'(\varphi_{\mb p}(t)) > 0, \,\, \forall t> \tau \},$$
$$\mathcal{B}_{-}= \{\mb  p \in \mathcal{B}\setminus W^{ss}(\mb p^*) \,| \,\,\, y_{m}'(\varphi_{\mb p}(t)) < 0, \,\, \forall t> \tau\},$$
The set $\mathcal{B}_{+}$ corresponds to the points whose orbits converge in the direction of positive growth of the infectious state $y_m$, and $\mathcal{B}_{-}$ in the descending direction.
\end{definition}

With this definition, we obtain the following result on the asymptotic behavior of the time average rate for system (\ref{sistema_general_1}).

\begin{theorem}\label{teorema_n+mD_real}
    Let $\mb p^*$ be a hyperbolic attracting equilibrium of system (\ref{sistema_general_1}) and let 
    $${\rm Re}(\lambda_{d}) \leq {\rm Re}(\lambda_{d-1}) \leq ... \leq {\rm Re}(\lambda_2)< \lambda_{1} < 0 $$
     be the eigenvalues of $\mb p^*$.
  Let us  denote the leading  eigenvector as
\begin{equation*}
        \mb v_{1} = \left(% \begin{matrix}
a_1,
\ldots,
a_n,
b_1,
\ldots,
b_m
%\end{matrix}
 \right)^t,
    \end{equation*}
     with $\|{\mb v_1}\| = 1 $ and assume $b_m>0$.
     Let $\mb u$ be a unit vector orthogonal  to the disease-free hyperplane given by
 \begin{equation}\nonumber
 \mb u=(\underbrace{0,...,0}_\text{n},u_1,...,u_m),
 \end{equation}
where $\sum_{i=1}^mu_i^2=1$ and $u_i\geq0$, $i=1,\ldots, m$.
 If $\mb p_0 \in  \mathcal{B}_{+}$ then
   $$ \lim_{t \to \infty}   X^{\textbf{u}}_{\mb p_0}(t) =  u_1b_1 + ... + u_m b_m. $$
 If $p_0 \in  \mathcal{B}_{-}$ then
   $$ \lim_{t \to \infty}  X^{\textbf{u}}_{\mb p_0} (t) = -(u_1b_1 + ... + u_m b_m). $$
   \end{theorem}
  
In the previous theorem we obtain a result that, although it requires certain hypotheses, is quite general. In the event that the coefficient $b_m=0$, we have $y_{m}'(\varphi_{\mb p}(t))=0$, for $t\gg1$. Hence, the disease is neither increasing nor decreasing in the direction of $y_m$ as the solution $\varphi_{\mb p_0}(t)$ converges to $\mb p^*$. %Luego, no nos entrega información sobre el sentido en el cual la solución está convergiendo a $p^*$.
In such case, we can choose that component of $\mb v_1$, among those associated with disease states, that is different from 0, say $b_{m-1} \neq 0$. Thus, the value of $s$ in
\eqref{ec_tasa_tiempo_n+m} will now be given by the sign of the rate of change of the solution in the direction of $y_{m-1}$ and the sets $\mathcal{B}_{+}$ and $\mathcal{ B}_{-}$ may be redefined, respectively, as:
 $$\mathcal{B}_{+}= \{ \mb p \in \mathcal{B}\setminus W^{ss}(\mb p^*) \, |  \,\,\, y_{m-1}'(\varphi_{\mb p}(t)) > 0, \,\, \forall t> \tau  \},$$
$$\mathcal{B}_{-} = \{ \mb p \in \mathcal{B}\setminus W^{ss}(\mb p^*)\,|  \,\,\, y_{m-1}'(\varphi_{\mb p}(t)) < 0, \,\, \forall t> \tau  \}.$$

Thus, the result of Theorem~\ref{teorema_n+mD_real} can be stated in terms of any component of $\mb v_1$ such that $b_i > 0$, $i=1,...,m$, redefining the sets $\mathcal{B}_+ $ and $\mathcal{B}_-$ appropriately.

%%%%%%%%%%%%%%%%%%%%%%%%
\subsection{Example: Time evolution of COVID-19 pandemics since March 1, 2022 in Chile}
%%%%%%%%%%%%%%%%%%%%%%%%

We return to model \eqref{sistema_SVEAIR} with parameter values as in table~\ref{Tabla2} to illustrate the results in this section. 
Starting from the initial point~\eqref{punto_inicial_Chile}, we denote the solution of~\eqref{sistema_SVEAIR} as  $\varphi_{\mb p_0}(t)$ with $\varphi_{\mb p_0}(0) = \mb p_0$. Thus, given a unit vector $\textbf{u}$ orthogonal to the disease-free surface, consider the propagation rate in the $\textbf{u}$ direction as a function of time $X^{\textbf{u}}_{\mb p_{0}}(t) $ defined in~\eqref{eq:tasa_tiempo}.
%
%\begin{equation}\label{Chile_tasa_tiempo}
 %   X^{\textbf{u}}_{\mb p_{0}}(t) :=   X^{\textbf{u}}(\varphi_{\mb p_0}(t)), \: \:\: 0\leq t \leq T.
%\end{equation}
%
Integrating the system numerically it is possible to calculate and plot the rate \eqref{eq:tasa_tiempo} over a time interval of duration $T>0$.  Considering an equispaced time discretization
$t_{k+1} - t_{k} = \frac{T}{\tau}, \:\:\: k \in \{0,...,\tau\}$ from $t_0 = 0$ to $t_{\tau} = T$ we obtain
\begin{equation}\label{Chile_tasa_discreta}
 X^{\textbf{u}}_{\mb p_{0}}(t_k) =   X^{\textbf{u}}(\varphi_{\mb p_0}(t_k)),  \:\:\: k \in \{0,...,\tau\}.
\end{equation}
%
%thus getting the expected behavior of the propagation rate in a period of $T$ days.

%Ahora, tomamos un  periodo de tiempo  de $T=60$ días, con $\tau = 60$, es decir, un espaciado de $T/\tau = 1$ día.  Con el solver $ode45$ de \textsc{Matlab} y los valores de parámetros de la tabla \ref{Tabla2}, calculamos numéricamente la solución del sistema desde el punto \eqref{punto_inicial_Chile} y con eso las tasas \eqref{Chile_tasa_discreta} tomando $\textbf{u}$ en dirección a los expuestos, infectados, asintomáticos y uniforme, tasas ya definidas en \eqref{sistema1_tasa_expuestos}, \eqref{sistema1_tasa_infectados}, \eqref{sistema1_tasa_asintomática}  y \eqref{sistema1_tasa_unif}, respectivamente. 

\begin{figure}[h!]
\centering
%\subfloat[]{\label{Chile_tiempo_expuestos}\includegraphics[scale=0.395]{Imagenes/Tasa_tiempo_expuestos.png}}
%\subfloat[]{\label{Chile_tiempo_infectados}\includegraphics[scale=0.395]{Imagenes/Tasa_tiempo_Infectados.png}}\\
%\subfloat[]{\label{Chile_tiempo_asintomaticos}\includegraphics[scale=0.395]{Imagenes/Tasa_tiempo_asintomatico.png}}
%\subfloat[]{\label{Chile_tiempo_uniforme}\includegraphics[scale=0.395]{Imagenes/Tasa_tiempo_uniforme.png}}
    \includegraphics[width=\textwidth]{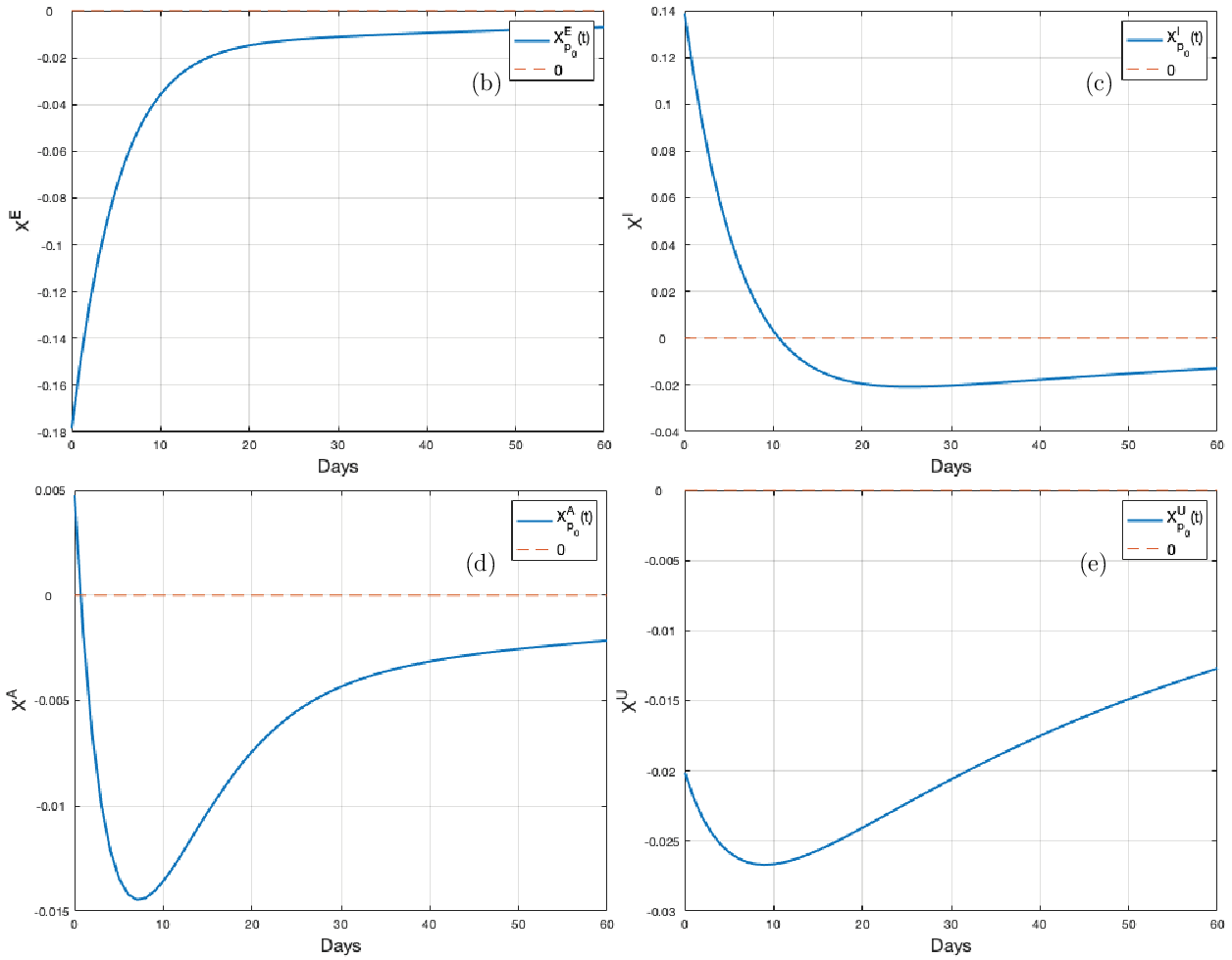}
\caption{
Plots of propagation rates over time, $X^{\textbf{u}}_{\mb p_{0}}(t)$, from the point $\mb p_0$ defined in \eqref{punto_inicial_Chile} with the different directions associated with disease states taken from model \eqref{sistema_SVEAIR}: (a)  In the direction of the exposed, $X^{E}_{\mb p_{0}}(t)$; (b) In the direction of symptomatic infected, $X^{I}_{\mb p_{0}}(t)$;  (c) In the direction of the asymtpmatic, $X^{A}_{\mb p_{0}}(t)$; (d) In the direction of uniform growth towards all the disease compartments, $X^{U}_{\mb p_{0}}(t)$. The model parameters are fixed according to the values in table~\ref{Tabla2}.}
\label{Chile_tiempo_EIAU}
\end{figure}

Figure \ref{Chile_tiempo_EIAU} shows the propagation rates \eqref{sistema1_tasa_expuestos}, \eqref{sistema1_tasa_infectados}, \eqref{sistema1_tasa_asintomática} and \eqref{sistema1_tasa_unif} of COVID-19 in Chile for a period of $T=60$ days after March 1, 2022.
%Recordar que hasta aquí conocíamos solo los valores de las tasas de propagación para esa fecha en particular, ver \eqref{tasas_2marzo2022}. De la figura \ref{Chile_tiempo_expuestos}
The numerical evidence indicates that the rate $X^E$ grows in time while remaining negative. On the other hand, the propagation rate of the symptomatic infected $X^I$ declines during the first 10 days until it becomes negative. 
Further, the propagation rate of the asymptomatic $X^A$ decreases until reaching 0 in the first 3 days, then it continues decreasing and reaches a minimum around day 8 and then it increases again but remains in negative values until day 60. Finally, considering all three aggregated disease compartments
 we observe that $X^U$ remains negative during the 60 days from March 1, 2022 and that COVID-19 continues to decrease over that period.

Using \eqref{Chile_tasa_discreta} we calculate the time average rate defined in \eqref{eq:promediot}  as:
\begin{equation}\label{aprox_tasa_promedio_temporal}
  \langle X_{\mb p_0}^{\textbf{u}} \rangle_{T}  \approx \frac{1}{\tau +1 } \sum_{k=0}^{T} X^{\textbf{u}}_{\mb p_{0}}(t_k).
\end{equation}
This makes it possible to determine the average growth tendency over $60$ days, obtaining:
\begin{equation}\label{promedio_temporal_60}
    \begin{array}{cc}
       \langle X_{\mb p_0}^{E} \rangle_{60} = -0.026, \ 
       \langle X_{\mb p_0}^{I} \rangle_{60} = -0.0036, \vspace{2mm} \\
        \langle X_{\mb p_0}^{A} \rangle_{60} = -0.0058, \ 
         \langle X_{\mb p_0}^{U} \rangle_{60} = -0.02.
   \end{array}
\end{equation}
In other words, there was a downward trend on the number of infected people in Chile in the 60 days from March 1, 2022.
%Now, we could ask ourselves, is this extrapolation consistent with the historical data for March and April 2022 in Chile?
The result~\eqref{promedio_temporal_60} agrees with the historical data taken from official sources~\cite{pasoapaso}, namely, COVID-19 in Chile was indeed declining in that same period of time. While our extrapolated results do not take into account actual changes in parameter values or in health policies in that period of time (variations in vaccination rates, contact rates, parameters associated with entry policies, etc.), in retrospect it manages to give an accurate qualitative picture of what was observed in reality.

%Suponiendo que los valores de parámetros se mantienen fijos por un largo periodo de tiempo, podríamos preguntarnos ¿qué se puede esperar en el largo plazo ($T\to\infty$) sobre las tasas promedio temporal? ó análogamente, ¿cuál será el valor esperado para las tasas de propagación a largo plazo?. Para responder esta pregunta utilizamos los resultados teóricos de la seccion XX. Allí demostramos que los valores asintóticos dependen de los valores propios dominantes asociados al punto de equilibrio asintóticamente estable existente.  Concluimos que, si el valor propio dominante es real el comportamiento asintótico de $\langle X^{\textbf{u}}_{p_0}\rangle_T $ depende de las componentes del vector propio principal. Por otro lado, si los valores propios dominantes son complejos conjugados, obtuvimos que $\lim_{T\to\infty} \langle X^{\textbf{u}}_{p_0}\rangle_T = 0$.

%\begin{figure}[h]
%\centering
%\includegraphics[scale=0.6]{Imagenes/Valores_Propios.png}
%    \includegraphics[scale=0.6]{figuras_eps/Valores_Propios.eps}
%\caption{Representation of the eigenvalues associated with the Jacobian matrix of \eqref{sistema_SVEAIR} at the equilibrium \eqref{punto_equilibrio_chile}. The leading eigenvalue is real. The model parameters are fixed according to the values in the table~\ref{Tabla2}.}
%\label{Chile_valores_propios}
%\end{figure}
%

 In terms of the asymptotic behavior, system \eqref{sistema_SVEAIR} has an endemic equilibrium point at
 %
%\begin{equation}\label{punto_equilibrio_chile}
   $  \mb p^*=(S^*,V^*,R^*,E^*,I^*,A^*) \approx (1.393.565, 515.344,131.264, 12.692, 18.964 ,3.712).$
% \end{equation}
 %
 %The eigenvalues of $\mb p^*$ are plotted in the complex plane in figure~\ref{Chile_valores_propios}.
 The leading eigenvalue is $\lambda_1\approx-0.007$.
 % it follows that $\mb p^*$ is a hyperbolic attractor and the leading eigenvalue is real. %Hence, we can use  theorem~\ref{teorema_n+mD_real}.  
 The associated leading eigenvector is
 $$\mb v_1 \approx (
0.950469,
0.309895,
0.023679,
0.002037,
0.002994,
0.000563
)^t$$
and we choose a unit vector 
 $\textbf{u} = \frac{1}{\sqrt{3}} (0,0,0,1,1,1)$
 pointing uniformly in direction to the three disease-associated compartments.  A numerical approximation 
 reveals that $\varphi_{\mb p_0}(t)$ approaches $\mb p^*$ in the direction of $-\mb v_1$  ---actually, for decreasing values of all three disease variables; compare with the negative values of every propagation rate in figure~\ref{Chile_tiempo_EIAU} for $t\gg1$. Hence, $s=-1$ and 
 $\mb p_{0} \in \mathcal{B}_{-}$. Therefore, from theorem~\ref{teorema_n+mD_real} we have
$$ \lim_{T \to \infty}    X^{\textbf{u}}_{\mb p_0}(t)  = - \frac{1}{\sqrt{3}}(0.002037+
0.002994+
0.000563) = - 0.003229.$$
In conclusion, assuming that all parameter values in \eqref{sistema_SVEAIR} remain fixed, the disease in the long term will tend to decrease in the direction to the three states associated with the disease, although at a relatively low speed.

%%%%%%%%%%%%%%%%%%%%%%%%%%%%%%%%%%%%%%%
\section{Discussion}
\label{sec:discussion}
%%%%%%%%%%%%%%%%%%%%%%%%%%%%%%%%%%%%%%%

In this work we constructed a set of new epidemiological thresholds and assessed their value in addressing the general problems of spreading and containment of a disease with influx of infected individuals.
 These propagation rates make it possible to measure whether, at a certain instant in time, the disease is spreading or declining in the direction of one (or several) of the states associated with the disease. Hence, this family of new predictive indices is able to quantify the severity of an immigration of infectious individuals into a community, and identify the key parameters that are capable of changing or reversing the spread of a disease in specific mathematical models. 

%En la seccion xx demostramos la construcción de este índice, sus propiedades, los diferentes tipos de tasas de propagación según la dirección del vector unitario normal a la superficie libre de enfermedad y definimos dos tasas a partir de la original: la tasa promedio ortogonal y la tasa promedio temporal.\\

We obtained analytical results on the asymptotic behavior of the propagation rates in general multidimensional models. If the leading eigenvalue associated with the attracting equilibrium state is real, the propagation rate and the long-term time average rate depend solely on the following ingredients: The initial point (in a neighborhood of the equilibrium), the leading eigenvector, and the unit vector that defines the propagation direction of interest. The case of complex leading eigenvalues is the subject of current research by one of the authors.

%En XXX, calculamos el comportamiento asintótico de la \textit{tasa promedio temporal} para modelos epidemiológicos de 2 y 3 compartimentos para luego obtener resultados para un modelo general multidimensional. Para ello, utilizamos la teoría cualitativa de los sistemas dinámicos en una vecindad del punto de equilibrio existente (hiperbólico y asintóticamente estable) y obtuvimos que si el valor propio principal asociado a la matriz jacobiana del punto de equilibrio es real, la tasa de propagación y la tasa promedio temporal a largo plazo va a  depender únicamente del punto inicial (en una vecindad del punto de equilibrio), del vector propio principal del equilibrio y del vector unitario que define la dirección de propagación. Mientras que si el valor propio principal es complejo, la tasa promedio temporal a largo plazo tiende a 0. \\

We illustrated our results and the functionality of the new propagation rates with a 6-dimensional compartmental model of COVID-19 with immigration of infected individuals taken from~\citep{Modelo_inmigracion}. For an arbitrary initial point, we calculated the rate of spread in the direction of growth of the compartments of exposed ($E$), infected symptomatic ($I$), infected asymptomatic ($A$), and in a uniform direction to the three aggregated disease compartments. We also obtained, for each case, the threshold set that defines whether the disease is spreading or decreasing in each direction.
Using an analogy with the basic reproduction number $\mathcal{R}_0$, we took as initial state the entire susceptible population except for one individual (patient zero) located in one of the states associated with the disease; with this, we obtained a set of threshold indices $\mathcal{R}_{E_0}$, $\mathcal{R}_{I_0} $ and $\mathcal{R}_{A_0}$, where the subscript indicates the compartmental state the patient zero is in.
These indices have a threshold value of 1 and indicate whether the disease (considering the three disease states) is spreading or decreasing at the initial instant. 
The numbers $\mathcal{R}_{E_0}$, $\mathcal{R}_{I_0} $ and $\mathcal{R}_{A_0}$ are useful in that their interpretation as thresholds is similar to that of the ``classic" basic reproduction number $\mc R_0$. Hence, their evaluation and explanation come about as simple and practical for those already familiar with the ``classic" $\mc R_0$. 
Unlike $\mc R_0$, however, 
$\mathcal{R}_{E_0}$, $\mathcal{R}_{I_0} $ and $\mathcal{R}_{A_0}$ do not give information about the average number of secondary infections produced by the primary one or the speed at which the disease is spreading. Nevertheless, some of these details are readily provided by the magnitude of $X^{\mb u}(\mb{p_0})$. Hence, indices $\mathcal{R}_{E_0}$, $\mathcal{R}_{I_0} $, $\mathcal{R}_{A_0}$ and $X^{\mb u}(\mb{p_0})$ may complement each other to gain valuable insight on the dynamical progress of the disease.

Unlike the traditional $\mathcal{R}_{0}$, the main advantage of these proposed new indices is that they are thresholds that can be used with or without immigration of infected people into the system. All in all, the analytical results allow us to pinpoint the key model parameters and associated threshold conditions that favor/prevent an epidemic outbreak in a community with “open borders”.
Taking data from the relevant sources and fitting it to the model, we carried out a case study on the spread of COVID-19 in Chile. We first obtained the rates of spread of the disease upon the appearance of patient zero (March 2020) and concluded that quarantine was not an effective measure to reduce the speed of spread at that initial instant. Our indices and rates are consistent with the events at the time, namely, they indicate that the country was in the verge of being hit by the pandemic.
An interesting observation is that if there had been a hypothetical vaccination rate of $v=0.0036$ the aggregated propagation rate would have been cut in half. Secondly, we studied the COVID-19 scenario in Chile two years later (March 2022).
We investigated how the aggregated uniform propagation rate is affected by changes associated with border measures, contagion prevention measures, and the parameters of the strain at that moment.
Among the results, we obtained that to request a negative PCR is more efficient than a vaccination certificate at the border. Our findings also indicate that relaxing capacity restrictions in places of great influx of people could have caused the rate to become positive. Moreover, we showed that if the recovery time is reduced by 7 days, the rate of decrease of the disease would increase by more than $50\%$.
Finally, we calculated the time average rates by letting the system evolve for 60 days from March 1, 2002, obtaining that (on average) the disease tends to decrease in that period of time. This result agrees with the real data. For the long-term behavior of the propagation rate, we used the analytical results from section~\ref{sec:asymptotic}, concluding that (if model parameters remain fixed and without external human interventions) the disease will tend to decrease in the uniform direction to the three aggregated disease states.

In short, we proposed a collection of indices that do not require any initial hypothesis about the model. This advantage allows our scheme to be implemented in multiple epidemiological scenarios, diseases, or contexts, making it a very versatile tool. In particular, it is applicable in epidemiological models with immigration of infected individuals, transforming it into a novel tool and (as far as we know) unique of its kind.  In this way, our approach allows us to isolate and identify common underlying mathematical ingredients in concrete systems with the aim of establishing sufficient conditions for the spread or containment of a particular disease. Consequently, knowledge from this type of investigation emerges as a natural, valuable input to
propose and evaluate decision strategies aimed at disease prevention and control so that the relevant
decision-makers
adopt timely health policies in epidemic scenarios.
Our indices and propagation rates also showed to have a good predictive potential on the overall behavior of the COVID-19 model in the Chilean case.
However, one must be careful in that our propagation rates provide information on the spread of the disease at an instantaneous level only. Nevertheless, one can readily evaluate them at any given time, varying parameters and/or evaluation point, thus being able to obtain a reasonable qualitative picture of the spread of the disease in the long term.
Furthermore, our methods are flexible enough to be applied in models not necessarily related to diseases in which we are interested in evaluating the growth/decrease of an aggregation of state variables.

From here a wide range of investigations can be opened around relevant information that can be obtained from the propagation rates. For instance, to analyze how the evolution of a propagation rate changes under parameter variation in non autonomous systems. 
One may also be interested in study the propagation rates in models that include compartments related to lifestyle (population that eats healthy, does sports, etc.). How does a propagation rate change in that compartment versus that of the unhealthy ones? How would the number of infected or deaths change if the parameters associated with the healthy population compartment are increased? 
Another challenging aspect is to design and study asymptotic propagation indices of a given disease in settings when the endemic equilibrium state may undergo bifurcations. Indeed, as explained earlier, our methods are broad enough to be generalizable to scenarios where the disease is in later stages of progress.
Hence, one may investigate the consequences of qualitative transitions in the dynamics ---triggered by critical perturbations on the model parameters--- on the asymptotic behavior of a set of disease variables (with and without immigration).

\bmhead{Acknowledgments}

SG and PA thank Proyecto Interno UTFSM PILI1906. PA and IF also acknowledge Proyecto Basal CMM-UChile.

%Please refer to Journal-level guidance for any specific requirements.

\section*{Declarations}

%Some journals require declarations to be submitted in a standardised format. Please check the Instructions for Authors of the journal to which you are submitting to see if you need to complete this section. If yes, your manuscript must contain the following sections under the heading `Declarations':

\begin{itemize}
\item The authors declare that they have no conflict of interest.
%\item Ethics approval 
%\item Consent for publication
\item This manuscript has no associated data.
%\item Authors' contributions
\end{itemize}

%\noindent
%If any of the sections are not relevant to your manuscript, please include the heading and write `Not applicable' for that section. 

%\bigskip

%\bigskip\noindent
%Springer journals and proceedings: \url{https://www.springer.com/gp/editorial-policies}

%\bigskip\noindent
%Nature Portfolio journals: \url{https://www.nature.com/nature-research/editorial-policies}

%\bigskip\noindent
%\textit{Scientific Reports}: \url{https://www.nature.com/srep/journal-policies/editorial-policies}

%\bigskip\noindent
%BMC journals: \url{https://www.biomedcentral.com/getpublished/editorial-policies}
%\end{flushleft}

%%%%%%%%%%%%%%%%%%%%%%%%%%%%
\section{Appendix A}
\label{sec:moreR0}
%%%%%%%%%%%%%%%%%%

Here we introduce an approach to define thresholds similar to \eqref{R_I0} and \eqref{R_A0} obtained from expression \eqref{eq:v} for general compartmental models of the form \eqref{sistema_general_1}. 

In many cases direct application of the implicit function theorem on equation $X^{\mb u}(\mb{p})=0$ allows one to define a quantity $\mc R(\mb u,\mb{p})$ such that
$$\mc R(\mb u,\mb{p})<1 \hspace{4mm} \Leftrightarrow \hspace{4mm} X^{\mb u}(\mb{p})<0 \hspace{5mm} {\rm and} \hspace{5mm} \mc R(\mb u,\mb{p})>1 \hspace{4mm} \Leftrightarrow \hspace{4mm} X^{\mb u}(\mb{p})>0.$$
Therefore, the disease at $\mb{p}$ is decaying  in the direction of $\mb u$ if $\mc R(\mb u,\mb{p})<1$ and it is spreading if  $\mc R(\mb u,\mb{p})>1$. Thus $\mc R(\mb u,\mb{p})$ emerges as an epidemiological threshold for the model \eqref{sistema_general_1} with immigration of infectives. To be more precise, if $a_k\big(\Pi p_{n+k} + g_{k}(\mb p)\big)\neq0$ for some fixed $k\in\{1,\ldots,m\}$, then
 we may define
$$\mc R(\mb u,\mb{p})=\frac{\displaystyle-\sum_{\displaystyle i=1 \atop \displaystyle i\neq k}^mu_{i} \big(\Pi\ p_{n+i} +\: g_{i}(\mb p)\big)}{u_k\big(\Pi p_{n+k} + g_{k}(\mb p)\big)}$$
as a {\bf propagation threshold} of the disease. 

When evaluated at the initial point $\mb{p_0}$ the interpretation of $\mc R(\mb u,\mb{p_0})$ has a correspondence with that of the basic reproduction number $\mc{R}_0$. For $t>0$, $\mc R(\mb u,\mb{p})$ recalls the effective reproduction number (although they are built from, and account for, different conceptual settings). More precisely, if $\mc R(\mb u,\mb{p})<1$ (respectively, $>1$), then at the moment when the distribution of the population in the different compartments is given by $\mb{p}$, the disease is decreasing (respectively, expanding) in the direction of $\mb u$.

%%%%%%%%%%%%%%%%%%%%%%

\bibliography{sofia-jmb}% common bib file


\begin{thebibliography}{38}
\ifx \bisbn   \undefined \def \bisbn  #1{ISBN #1}\fi
\ifx \binits  \undefined \def \binits#1{#1}\fi
\ifx \bauthor  \undefined \def \bauthor#1{#1}\fi
\ifx \batitle  \undefined \def \batitle#1{#1}\fi
\ifx \bjtitle  \undefined \def \bjtitle#1{#1}\fi
\ifx \bvolume  \undefined \def \bvolume#1{\textbf{#1}}\fi
\ifx \byear  \undefined \def \byear#1{#1}\fi
\ifx \bissue  \undefined \def \bissue#1{#1}\fi
\ifx \bfpage  \undefined \def \bfpage#1{#1}\fi
\ifx \blpage  \undefined \def \blpage #1{#1}\fi
\ifx \burl  \undefined \def \burl#1{\textsf{#1}}\fi
\ifx \doiurl  \undefined \def \doiurl#1{\url{https://doi.org/#1}}\fi
\ifx \betal  \undefined \def \betal{\textit{et al.}}\fi
\ifx \binstitute  \undefined \def \binstitute#1{#1}\fi
\ifx \binstitutionaled  \undefined \def \binstitutionaled#1{#1}\fi
\ifx \bctitle  \undefined \def \bctitle#1{#1}\fi
\ifx \beditor  \undefined \def \beditor#1{#1}\fi
\ifx \bpublisher  \undefined \def \bpublisher#1{#1}\fi
\ifx \bbtitle  \undefined \def \bbtitle#1{#1}\fi
\ifx \bedition  \undefined \def \bedition#1{#1}\fi
\ifx \bseriesno  \undefined \def \bseriesno#1{#1}\fi
\ifx \blocation  \undefined \def \blocation#1{#1}\fi
\ifx \bsertitle  \undefined \def \bsertitle#1{#1}\fi
\ifx \bsnm \undefined \def \bsnm#1{#1}\fi
\ifx \bsuffix \undefined \def \bsuffix#1{#1}\fi
\ifx \bparticle \undefined \def \bparticle#1{#1}\fi
\ifx \barticle \undefined \def \barticle#1{#1}\fi
\bibcommenthead
\ifx \bconfdate \undefined \def \bconfdate #1{#1}\fi
\ifx \botherref \undefined \def \botherref #1{#1}\fi
\ifx \url \undefined \def \url#1{\textsf{#1}}\fi
\ifx \bchapter \undefined \def \bchapter#1{#1}\fi
\ifx \bbook \undefined \def \bbook#1{#1}\fi
\ifx \bcomment \undefined \def \bcomment#1{#1}\fi
\ifx \oauthor \undefined \def \oauthor#1{#1}\fi
\ifx \citeauthoryear \undefined \def \citeauthoryear#1{#1}\fi
\ifx \endbibitem  \undefined \def \endbibitem {}\fi
\ifx \bconflocation  \undefined \def \bconflocation#1{#1}\fi
\ifx \arxivurl  \undefined \def \arxivurl#1{\textsf{#1}}\fi
\csname PreBibitemsHook\endcsname

%%% 1
\bibitem[\protect\citeauthoryear{Brauer et~al.}{2012}]{brauer-cc}
\begin{bbook}
\bauthor{\bsnm{Brauer}, \binits{F.}},
\bauthor{\bsnm{Castillo-Chavez}, \binits{C.}},
\bauthor{\bsnm{Castillo-Chavez}, \binits{C.}}:
\bbtitle{Mathematical Models in Population Biology and Epidemiology}
vol. \bseriesno{2}.
\bpublisher{Springer},
\blocation{New York}
(\byear{2012})
\end{bbook}
\endbibitem

%%% 2
\bibitem[\protect\citeauthoryear{Delamater et~al.}{2019}]{delamater}
\begin{barticle}
\bauthor{\bsnm{Delamater}, \binits{P.L.}},
\bauthor{\bsnm{Street}, \binits{E.J.}},
\bauthor{\bsnm{Leslie}, \binits{T.F.}},
\bauthor{\bsnm{Yang}, \binits{Y.T.}},
\bauthor{\bsnm{Jacobsen}, \binits{K.H.}}:
\batitle{Complexity of the basic reproduction number ($\mathcal{R}_0$)}.
\bjtitle{Emerging Infectious Diseases}
\bvolume{25}(\bissue{1}),
\bfpage{1}
(\byear{2019})
\end{barticle}
\endbibitem

%%% 3
\bibitem[\protect\citeauthoryear{Diekmann and
  Heesterbeek}{2000}]{diekmann-book}
\begin{bbook}
\bauthor{\bsnm{Diekmann}, \binits{O.}},
\bauthor{\bsnm{Heesterbeek}, \binits{J.A.P.}}:
\bbtitle{Mathematical Epidemiology of Infectious Diseases: Model Building,
  Analysis and Interpretation}
vol. \bseriesno{5}.
\bpublisher{John Wiley \& Sons},
\blocation{Chichester}
(\byear{2000})
\end{bbook}
\endbibitem

%%% 4
\bibitem[\protect\citeauthoryear{Murray}{2002}]{murray}
\begin{bbook}
\bauthor{\bsnm{Murray}, \binits{J.D.}}:
\bbtitle{Mathematical Biology: I. An Introduction}.
\bpublisher{Springer},
\blocation{Berlin Heidelberg}
(\byear{2002})
\end{bbook}
\endbibitem

%%% 5
\bibitem[\protect\citeauthoryear{Smith}{2008}]{smith}
\begin{bbook}
\bauthor{\bsnm{Smith}, \binits{R.}}:
\bbtitle{Modelling Disease Ecology with Mathematics}.
\bpublisher{American Institute of Mathematical Sciences},
\blocation{Springfield}
(\byear{2008})
\end{bbook}
\endbibitem

%%% 6
\bibitem[\protect\citeauthoryear{Diekmann et~al.}{1990}]{diekmann}
\begin{barticle}
\bauthor{\bsnm{Diekmann}, \binits{O.}},
\bauthor{\bsnm{Heesterbeek}, \binits{J.A.P.}},
\bauthor{\bsnm{Metz}, \binits{J.A.}}:
\batitle{On the definition and the computation of the basic reproduction ratio
  r 0 in models for infectious diseases in heterogeneous populations}.
\bjtitle{Journal of Mathematical Biology}
\bvolume{28},
\bfpage{365}--\blpage{382}
(\byear{1990})
\end{barticle}
\endbibitem

%%% 7
\bibitem[\protect\citeauthoryear{Diekmann et~al.}{2010}]{MPG}
\begin{barticle}
\bauthor{\bsnm{Diekmann}, \binits{O.}},
\bauthor{\bsnm{Heesterbeek}, \binits{J.}},
\bauthor{\bsnm{Roberts}, \binits{M.G.}}:
\batitle{The construction of next-generation matrices for compartmental
  epidemic models}.
\bjtitle{Journal of the Royal Society Interface}
\bvolume{7}(\bissue{47}),
\bfpage{873}--\blpage{885}
(\byear{2010})
\end{barticle}
\endbibitem

%%% 8
\bibitem[\protect\citeauthoryear{Van~den Driessche and Watmough}{2002}]{vander}
\begin{barticle}
\bauthor{\bsnm{Driessche}, \binits{P.}},
\bauthor{\bsnm{Watmough}, \binits{J.}}:
\batitle{Reproduction numbers and sub-threshold endemic equilibria for
  compartmental models of disease transmission}.
\bjtitle{Mathematical Biosciences}
\bvolume{180}(\bissue{1-2}),
\bfpage{29}--\blpage{48}
(\byear{2002})
\end{barticle}
\endbibitem

%%% 9
\bibitem[\protect\citeauthoryear{Brauer and van~den Driessche}{2001}]{Brauer}
\begin{barticle}
\bauthor{\bsnm{Brauer}, \binits{F.}},
\bauthor{\bsnm{Driessche}, \binits{P.}}:
\batitle{Models for transmission of disease with immigration of infectives}.
\bjtitle{Mathematical Biosciences}
\bvolume{171}(\bissue{2}),
\bfpage{143}--\blpage{154}
(\byear{2001})
\end{barticle}
\endbibitem

%%% 10
\bibitem[\protect\citeauthoryear{Guo and Li}{2012}]{guo12}
\begin{barticle}
\bauthor{\bsnm{Guo}, \binits{H.}},
\bauthor{\bsnm{Li}, \binits{M.Y.}}:
\batitle{Impacts of migration and immigration on disease transmission dynamics
  in heterogeneous populations}.
\bjtitle{Discrete Contin. Dyn. Syst. Ser. B}
\bvolume{17}(\bissue{7}),
\bfpage{2413}--\blpage{2430}
(\byear{2012})
\end{barticle}
\endbibitem

%%% 11
\bibitem[\protect\citeauthoryear{Guo and Wu}{2011}]{guo11}
\begin{barticle}
\bauthor{\bsnm{Guo}, \binits{H.}},
\bauthor{\bsnm{Wu}, \binits{J.}}:
\batitle{Persistent high incidence of tuberculosis among immigrants in a
  low-incidence country: impact of immigrants with early or late latency}.
\bjtitle{Math Biosci Eng}
\bvolume{8}(\bissue{3}),
\bfpage{695}--\blpage{709}
(\byear{2011})
\end{barticle}
\endbibitem

%%% 12
\bibitem[\protect\citeauthoryear{Henshaw and McCluskey}{2015}]{PE_1}
\begin{barticle}
\bauthor{\bsnm{Henshaw}, \binits{S.}},
\bauthor{\bsnm{McCluskey}, \binits{C.C.}}:
\batitle{Global stability of a vaccination model with immigration}.
\bjtitle{Electronic Journal of Differential Equations}
\bvolume{92},
\bfpage{1}--\blpage{10}
(\byear{2015})
\end{barticle}
\endbibitem

%%% 13
\bibitem[\protect\citeauthoryear{Jia et~al.}{2008}]{Modelo_3}
\begin{barticle}
\bauthor{\bsnm{Jia}, \binits{Z.-W.}},
\bauthor{\bsnm{Tang}, \binits{G.-Y.}},
\bauthor{\bsnm{Jin}, \binits{Z.}},
\bauthor{\bsnm{Dye}, \binits{C.}},
\bauthor{\bsnm{Vlas}, \binits{S.J.}},
\bauthor{\bsnm{Li}, \binits{X.-W.}},
\bauthor{\bsnm{Feng}, \binits{D.}},
\bauthor{\bsnm{Fang}, \binits{L.-Q.}},
\bauthor{\bsnm{Zhao}, \binits{W.-J.}},
\bauthor{\bsnm{Cao}, \binits{W.-C.}}:
\batitle{Modeling the impact of immigration on the epidemiology of
  tuberculosis}.
\bjtitle{Theoretical Population Biology}
\bvolume{73}(\bissue{3}),
\bfpage{437}--\blpage{448}
(\byear{2008})
\end{barticle}
\endbibitem

%%% 14
\bibitem[\protect\citeauthoryear{Naresh et~al.}{2009}]{naresh}
\begin{barticle}
\bauthor{\bsnm{Naresh}, \binits{R.}},
\bauthor{\bsnm{Tripathi}, \binits{A.}},
\bauthor{\bsnm{Sharma}, \binits{D.}}:
\batitle{Modelling and analysis of the spread of aids epidemic with immigration
  of hiv infectives}.
\bjtitle{Mathematical and Computer Modelling}
\bvolume{49}(\bissue{5-6}),
\bfpage{880}--\blpage{892}
(\byear{2009})
\end{barticle}
\endbibitem

%%% 15
\bibitem[\protect\citeauthoryear{Sigdel and McCluskey}{2014}]{PE_2}
\begin{barticle}
\bauthor{\bsnm{Sigdel}, \binits{R.P.}},
\bauthor{\bsnm{McCluskey}, \binits{C.C.}}:
\batitle{Global stability for an sei model of infectious disease with
  immigration}.
\bjtitle{Applied Mathematics and Computation}
\bvolume{243},
\bfpage{684}--\blpage{689}
(\byear{2014})
\end{barticle}
\endbibitem

%%% 16
\bibitem[\protect\citeauthoryear{Tumwiine et~al.}{2010}]{tumwiine}
\begin{barticle}
\bauthor{\bsnm{Tumwiine}, \binits{J.}},
\bauthor{\bsnm{Mugisha}, \binits{J.}},
\bauthor{\bsnm{Luboobi}, \binits{L.}}:
\batitle{A host-vector model for malaria with infective immigrants}.
\bjtitle{Journal of Mathematical Analysis and Applications}
\bvolume{361}(\bissue{1}),
\bfpage{139}--\blpage{149}
(\byear{2010})
\end{barticle}
\endbibitem

%%% 17
\bibitem[\protect\citeauthoryear{Almarashi and McCluskey}{2019}]{Almarashi}
\begin{barticle}
\bauthor{\bsnm{Almarashi}, \binits{R.M.}},
\bauthor{\bsnm{McCluskey}, \binits{C.C.}}:
\batitle{The effect of immigration of infectives on disease-free equilibria}.
\bjtitle{Journal of Mathematical Biology}
\bvolume{79},
\bfpage{1015}--\blpage{1028}
(\byear{2019})
\end{barticle}
\endbibitem

%%% 18
\bibitem[\protect\citeauthoryear{McCluskey}{2021}]{mccluskey}
\begin{botherref}
\oauthor{\bsnm{McCluskey}, \binits{C.}}:
Lyapunov functions for disease models with immigration of infected hosts.
Discrete \& Continuous Dynamical Systems-Series B
\textbf{26}(8)
(2021)
\end{botherref}
\endbibitem

%%% 19
\bibitem[\protect\citeauthoryear{McLure and Glass}{2020}]{mclure}
\begin{barticle}
\bauthor{\bsnm{McLure}, \binits{A.}},
\bauthor{\bsnm{Glass}, \binits{K.}}:
\batitle{Some simple rules for estimating reproduction numbers in the presence
  of reservoir exposure or imported cases}.
\bjtitle{Theoretical Population Biology}
\bvolume{134},
\bfpage{182}--\blpage{194}
(\byear{2020})
\end{barticle}
\endbibitem

%%% 20
\bibitem[\protect\citeauthoryear{Ayana et~al.}{2020}]{ayana}
\begin{barticle}
\bauthor{\bsnm{Ayana}, \binits{M.}},
\bauthor{\bsnm{Hailegiorgis}, \binits{T.}},
\bauthor{\bsnm{Getnet}, \binits{K.}}:
\batitle{The impact of infective immigrants and self isolation on the dynamics
  and spread of covid-19 pandemic: A mathematical modeling study}.
\bjtitle{Pure and Applied Mathematics Journal}
\bvolume{9}(\bissue{6}),
\bfpage{109}--\blpage{117}
(\byear{2020})
\end{barticle}
\endbibitem

%%% 21
\bibitem[\protect\citeauthoryear{Tchoumi et~al.}{2022}]{Modelo_inmigracion}
\begin{barticle}
\bauthor{\bsnm{Tchoumi}, \binits{S.Y.}},
\bauthor{\bsnm{Rwezaura}, \binits{H.}},
\bauthor{\bsnm{Diagne}, \binits{M.L.}},
\bauthor{\bsnm{Gonz{\'a}lez-Parra}, \binits{G.}},
\bauthor{\bsnm{Tchuenche}, \binits{J.}}:
\batitle{Impact of infective immigrants on covid-19 dynamics}.
\bjtitle{Mathematical and Computational Applications}
\bvolume{27}(\bissue{1}),
\bfpage{11}
(\byear{2022})
\end{barticle}
\endbibitem

%%% 22
\bibitem[\protect\citeauthoryear{Castillo-Laborde et~al.}{2021}]{carla2}
\begin{barticle}
\bauthor{\bsnm{Castillo-Laborde}, \binits{C.}},
\bauthor{\bsnm{Gajardo}, \binits{P.}},
\bauthor{\bsnm{Ferrari}, \binits{N.-D.}},
\bauthor{\bsnm{Matute}, \binits{I.}},
\bauthor{\bsnm{Hirmas-Adauy}, \binits{M.}},
\bauthor{\bsnm{Aguirre}, \binits{P.}},
\bauthor{\bsnm{Ram{\'\i}rez}, \binits{H.}},
\bauthor{\bsnm{Ram{\'\i}rez}, \binits{D.}},
\bauthor{\bsnm{Aguilera}, \binits{X.}}, \betal:
\batitle{Modelling cost-effectiveness of syphilis detection strategies in
  prisoners: exploratory exercise in a chilean male prison}.
\bjtitle{Cost Effectiveness and Resource Allocation}
\bvolume{19}(\bissue{1}),
\bfpage{1}--\blpage{9}
(\byear{2021})
\end{barticle}
\endbibitem

%%% 23
\bibitem[\protect\citeauthoryear{{Gobierno de Chile}}{}]{minsal}
\begin{botherref}
\oauthor{\bsnm{{Gobierno de Chile}}}:
Seguimos Cuid\'andonos. Informaci\'on sobre la situaci\'on epidemiol\'ogica de
  Covid-19 en Chile, el estado comunal, adem\'as de medidas y recomendaciones
  del Ministerio de Salud para prevenir esta enfermedad.
\url{https://www.gob.cl/pasoapaso/}.
Accessed: 2023-06-20
\end{botherref}
\endbibitem

%%% 24
\bibitem[\protect\citeauthoryear{{World Health Organization}}{}]{who2}
\begin{botherref}
\oauthor{\bsnm{{World Health Organization}}}:
Coronavirus disease (COVID-19) pandemic.
\url{https://www.who.int/emergencies/diseases/novel-coronavirus-2019}.
Accessed: 2023-06-20
\end{botherref}
\endbibitem

%%% 25
\bibitem[\protect\citeauthoryear{{United Nations}}{}]{poblacion_chile_2020}
\begin{botherref}
\oauthor{\bsnm{{United Nations}}}:
UN Population Division Data Portal, Interactive access to global demographic
  indicators.
\url{https://population.un.org/dataportal/home}.
Accessed: 2023-06-20
\end{botherref}
\endbibitem

%%% 26
\bibitem[\protect\citeauthoryear{{Aeropuerto de Santiago}}{}]{nuevo_pudahuel}
\begin{botherref}
\oauthor{\bsnm{{Aeropuerto de Santiago}}}:
Nuevo Pudahuel.
\url{https://www.nuevopudahuel.cl/fflights}.
Accessed: 2023-06-20
\end{botherref}
\endbibitem

%%% 27
\bibitem[\protect\citeauthoryear{FlightAware}{}]{flight_trakking}
\begin{botherref}
\oauthor{\bsnm{FlightAware}}:
Flight Finder.
\url{https://es.flightaware.com}.
Accessed: 2022-06-20
\end{botherref}
\endbibitem

%%% 28
\bibitem[\protect\citeauthoryear{{Expansi\'on/Datosmacro.com}}{}]{mortalidad_chile}
\begin{botherref}
\oauthor{\bsnm{{Expansi\'on/Datosmacro.com}}}:
Chile - Mortalidad.
\url{https://datosmacro.expansion.com/demografia/mortalidad/chile}.
Accessed: 2023-06-20
\end{botherref}
\endbibitem

%%% 29
\bibitem[\protect\citeauthoryear{{Ministerio de Salud, Gobierno de
  Chile}}{}]{casos_activos_Chile}
\begin{botherref}
\oauthor{\bsnm{{Ministerio de Salud, Gobierno de Chile}}}:
Cifras Oficiales COVID-19.
\url{https://www.minsal.cl/wp-content/uploads/2022/03/CP-REPORTE-COVID-19-Martes-01.03.2022.pdf}.
Accessed: 2023-06-20
\end{botherref}
\endbibitem

%%% 30
\bibitem[\protect\citeauthoryear{Oran and
  Topol}{2021}]{proporcion_asintomaticos}
\begin{barticle}
\bauthor{\bsnm{Oran}, \binits{D.P.}},
\bauthor{\bsnm{Topol}, \binits{E.J.}}:
\batitle{The proportion of sars-cov-2 infections that are asymptomatic: a
  systematic review}.
\bjtitle{Annals of Internal Medicine}
\bvolume{174}(\bissue{5}),
\bfpage{655}--\blpage{662}
(\byear{2021})
\end{barticle}
\endbibitem

%%% 31
\bibitem[\protect\citeauthoryear{{Gobierno de Chile}}{}]{pasoapaso}
\begin{botherref}
\oauthor{\bsnm{{Gobierno de Chile}}}:
Cifras Oficiales COVID-19.
\url{https://www.gob.cl/pasoapaso/cifrasoficiales/\#datos-c19}.
Accessed: 2023-06-20
\end{botherref}
\endbibitem

%%% 32
\bibitem[\protect\citeauthoryear{Widge et~al.}{2021}]{inmunidad_vacuna}
\begin{barticle}
\bauthor{\bsnm{Widge}, \binits{A.T.}},
\bauthor{\bsnm{Rouphael}, \binits{N.G.}},
\bauthor{\bsnm{Jackson}, \binits{L.A.}},
\bauthor{\bsnm{Anderson}, \binits{E.J.}},
\bauthor{\bsnm{Roberts}, \binits{P.C.}},
\bauthor{\bsnm{Makhene}, \binits{M.}},
\bauthor{\bsnm{Chappell}, \binits{J.D.}},
\bauthor{\bsnm{Denison}, \binits{M.R.}},
\bauthor{\bsnm{Stevens}, \binits{L.J.}},
\bauthor{\bsnm{Pruijssers}, \binits{A.J.}}, \betal:
\batitle{Durability of responses after sars-cov-2 mrna-1273 vaccination}.
\bjtitle{New England Journal of Medicine}
\bvolume{384}(\bissue{1}),
\bfpage{80}--\blpage{82}
(\byear{2021})
\end{barticle}
\endbibitem

%%% 33
\bibitem[\protect\citeauthoryear{{Expansi\'on/Datosmacro.com}}{}]{piramide_poblacion}
\begin{botherref}
\oauthor{\bsnm{{Expansi\'on/Datosmacro.com}}}:
Chile - Pir\'amide de poblaci\'on.
\url{https://datosmacro.expansion.com/demografia/estructura-poblacion/chile}.
Accessed: 2023-06-20
\end{botherref}
\endbibitem

%%% 34
\bibitem[\protect\citeauthoryear{Bever}{}]{inmunidad_vacuna_2}
\begin{botherref}
\oauthor{\bsnm{Bever}, \binits{L.}}:
How long will the coronavirus vaccines protect you? Experts weigh in.
\url{https://www.washingtonpost.com/lifestyle/2021/03/29/how-long-immunity-lasts-covid-vaccine/}.
Accessed: 2023-06-20
\end{botherref}
\endbibitem

%%% 35
\bibitem[\protect\citeauthoryear{{Consejo para la
  Transparencia}}{}]{portal_transparencia_Chile}
\begin{botherref}
\oauthor{\bsnm{{Consejo para la Transparencia}}}:
Portal Transparencia Chile.
\url{https://www.portaltransparencia.cl/PortalPdT/}.
Accessed: 2023-06-20
\end{botherref}
\endbibitem

%%% 36
\bibitem[\protect\citeauthoryear{{Imperial College
  London}}{}]{proporcion_expuestos_infectados}
\begin{botherref}
\oauthor{\bsnm{{Imperial College London}}}:
MRC Centre for Global Infectious Disease Analysis.
\url{https://www.imperial.ac.uk/mrc-global-infectious-disease-analysis/covid-19/}.
Accessed: 2023-06-20
\end{botherref}
\endbibitem

%%% 37
\bibitem[\protect\citeauthoryear{Guckenheimer and Holmes}{2013}]{GH}
\begin{bbook}
\bauthor{\bsnm{Guckenheimer}, \binits{J.}},
\bauthor{\bsnm{Holmes}, \binits{P.}}:
\bbtitle{Nonlinear Oscillations, Dynamical Systems, and Bifurcations of Vector
  Fields}
vol. \bseriesno{42}.
\bpublisher{Springer},
\blocation{New York}
(\byear{2013})
\end{bbook}
\endbibitem

%%% 38
\bibitem[\protect\citeauthoryear{Shilnikov
  et~al.}{1998}]{Teoria_cualitativa_sistemas}
\begin{bbook}
\bauthor{\bsnm{Shilnikov}, \binits{L.P.}},
\bauthor{\bsnm{Shilnikov}, \binits{A.L.}},
\bauthor{\bsnm{Turaev}, \binits{D.V.}},
\bauthor{\bsnm{Chua}, \binits{L.O.}}:
\bbtitle{Methods Of Qualitative Theory In Nonlinear Dynamics (Part I)}
vol. \bseriesno{4}.
\bpublisher{World Scientific},
\blocation{Singapore}
(\byear{1998})
\end{bbook}
\endbibitem

\end{thebibliography}

\end{document}